\documentclass[referee]{jfm}
\usepackage{natbib}
\usepackage{subfigure}
\usepackage{bm}
\usepackage{verbatim}

\newif\ifpdf
\ifx\pdfoutput\undefined
\pdffalse 
\else
\pdfoutput=1 
\pdftrue
\fi
\ifpdf
\usepackage[pdftex]{graphicx}
\else
\usepackage{graphicx}
\fi
\ifpdf
\DeclareGraphicsExtensions{.pdf, .jpg, .tif}
\else
\DeclareGraphicsExtensions{.eps, .jpg}
\fi

\ifCUPmtlplainloaded \else
  \checkfont{eurm10}
  \iffontfound
    \IfFileExists{upmath.sty}
      {\typeout{^^JFound AMS Euler Roman fonts on the system,
                   using the 'upmath' package.^^J}%
       \usepackage{upmath}}
      {\typeout{^^JFound AMS Euler Roman fonts on the system, but you
                   dont seem to have the}%
       \typeout{'upmath' package installed. JFM.cls can take advantage
                 of these fonts,^^Jif you use 'upmath' package.^^J}%
      }
  \else
  \fi
\fi

\ifCUPmtlplainloaded \else
  \checkfont{msam10}
  \iffontfound
    \IfFileExists{amssymb.sty}
      {\typeout{^^JFound AMS Symbol fonts on the system, using the
                'amssymb' package.^^J}%
       \usepackage{amssymb}%
         \let\leq=\leqslant
         
      }{}
  \fi
\fi

\ifCUPmtlplainloaded \else
  \IfFileExists{amsbsy.sty}
    {\typeout{^^JFound the 'amsbsy' package on the system, using it.^^J}%
     \usepackage{amsbsy}}
    {}
\fi

\newsavebox{\astrutbox}
\sbox{\astrutbox}{\rule[-5pt]{0pt}{20pt}}

\title[High-wavenumber energy spectra of rotating Boussinesq]{Scaling of high-wavenumber energy spectra in the unit aspect-ratio rotating Boussinesq system}

\author[S.~Kurien]{S\ls U\ls S\ls A\ls N\ns K\ls U\ls R\ls I\ls E\ls N}

\affiliation{Theoretical Division, Los Alamos National Laboratory, Los
  Alamos, NM 87545, USA}

\date{\today}

\begin{document}

\maketitle

\begin{abstract}Phenomenological and numerical studies of the small
  scale spectra of energy are presented for high Reynolds number
  rotating Boussinesq flows in unit aspect-ratio domains. We introduce
  a non-dimensional parameter $\Gamma({\bm k}) = \frac{f k_z}{N k_h}$,
  where the wavevector ${\bm k}$ has vertical component $k_z$ and
  horizontal component $k_h$, $f$ is the Coriolis frequency and $N$ is
  the Brunt-V\"ais\"al\"a frequency. While requiring that $f$ and $N$
  are such that the potential vorticity is nearly linear in the
  dynamical variables, we deduce that for $\Gamma \ll 1$, the
  potential enstrophy suppresses the transfer of horizontal kinetic
  energy into large $k_h$ modes while forcing it to become independent
  of $k_z$, scaling as $k_h^{-5}$.  For $\Gamma \gg 1$, the potential
  enstrophy suppresses the transfer of potential energy into the large
  $k_z$ modes while forcing it to become independent of $k_h$, scaling
  as $k_z^{-5}$.  The usual scaling exponent $-3$ is adjusted down to
  $-5$ based on a heuristic argument which links the downscale fluxes
  of potential enstrophy and energy by their ratio. Spectra computed
  from high-resolution simulations of the Boussinesq equations with
  isotropic low-wavenumber forcing are used to explore such
  anisotropic constraints on the energy and provide {\it a posteriori}
  justification for the joint flux ansatz used to obtain the $-5$
  scaling exponent. In simulations for which $f/N \ll 1$, the
  asymptotic regime of $\Gamma \ll 1$ is achieved and the horizontal
  kinetic energy is shown to become independent of $k_z$ and scale as
  $k_h^{-5}$. This collapse results in a uniform $k_z^{0}$ scaling for
  all $k_h$ and hence a uniform $k^{0}$ scaling of the energy in the
  spherical wavenumber, indicating extremely efficient net downscale
  transfer of energy.  In simulations for which $f/N \gg 1$, the trend
  towards the asymptotic regime of $\Gamma \gg 1$ is observed although
  the potential energy does not become entirely independent of $k_h$
  as predicted. In simulations with $f/N = 1$ where rotation and
  stratification are equally strong, both $\Gamma \ll 1$ and $\Gamma
  \gg 1$ regimes may be recovered depending on the aspect-ratio of the
  wavevector, $\tau = k_z/k_h$, which is also the cotangent of the
  wavevector angle to the vertical. The dependence of the
  spectra on $\tau$ results, on average, in shallower scaling exponent
  values between $-1$ and $-5/3$ in both $k_h$ and $k_z$ implying an
  effectively more isotropic downscale distribution of energy in this
  limit.  In all cases the empirical evidence points to both energy
  and potential enstrophy being jointly transferred downscale with the
  spectral scaling of the the former constrained by the latter.


\end{abstract}


\section{Introduction}
The Boussinesq approximation for hydrodynamic flow is the cornerstone
for many applications in which a suitable model for scalar transport
is needed. It is physically realistic in that it postulates that the
fluctuations of the scalar are small compared to its background value
and is computationally tenable for linear stratification profiles and
simple boundary conditions.  The Boussinesq approximation in the
rotating frame forms the parent system for the derivation of further
simplifications such as the shallow-water, hydrostatic and
quasi-geostrophic (QG) models which are standard approximations in
geophysical regimes which include rotation and stratification of the
flow.

The Boussinesq system for a stably stratified flow in a rotating frame
is given by:
\begin{eqnarray}
\frac{D}{Dt}\bm{u} + f\hat{\bm z}\times \bm{u} +
\nabla p + N\theta \hat{\bm{z}} &=& \nu\nabla^2 \bm{u} + {\cal F}\label{boussu}\\
\frac{D}{Dt}\theta - Nw &=& \kappa \nabla^2\theta \label{bousstta}\\
\nabla\cdot\bm{u} &=& 0,\nonumber\\
\displaystyle\frac{D}{Dt} &=& \frac{\partial}{\partial t} + \bm{u}\cdot\nabla, 
\end{eqnarray}
where $\bm{u}$ is the velocity, $w$ is its vertical
component, $p$ is the effective pressure and $\cal F$ is an external
input or force.  The total density is given by
${\cal \rho}_T(\bm{x}) = \rho_0 - bz + \rho(\bm{x}), \label{rho}$
such~that $\quad| \rho | \ll | bz | \ll \rho_0 $
where $\rho_0$ is the constant background, $b$ is constant and
larger than zero for stable stratification in the vertical
$z$-coordinate, $\rho$ is the density fluctuation. The density 
in normalized to $\displaystyle
\theta =\rho ({g}/{b \rho_0})^{1/2}$ which has the dimensions of velocity.
The Coriolis parameter $f = 2\Omega$ where $\Omega$ is the constant
rotation rate about the $z$-axis, the Brunt-V\"ais\"al\"a frequency
$\displaystyle N = ({gb}/{\rho_0})^{1/2}$, $\nu=\mu/\rho_0$ is the
kinematic viscosity and $\kappa$ is the mass diffusivity coefficient.
We assume periodic or infinite boundary conditions with Prandtl number $Pr = \nu/\kappa \sim {\cal O}(1)$.  The relevant
non-dimensional parameters for this system are the Rossby number $Ro =
f_{nl}/{f}$ and the Froude number $Fr = f_{nl}/{N}$, where $f_{nl} =
(\epsilon_f k_f^2)^{1/3}$ is the non-linear frequency given input rate
of energy $\epsilon_f$ \cite{SW02}. Thus $Ro$ and $Fr$ are the ratios
of rotation and stratification timescales respectively to the
nonlinear timescale.

There are three overlapping ideas from turbulence theory which we will
exploit in this analysis of the Boussinesq equations.  First from
\cite{K41b_5/3} the notion of a universal range of scales governed by
the dynamics of the flux of a conserved quantity.  Second, the
analysis of \cite{Kraichnan71b} on the inertial ranges of two-dimensional
flows based on the constraint imposed by the joint conservation of
both energy and enstrophy. Kraichnan thus deduced the $k^{-3}$ scaling,
up to logarithmic correction, of the energy in
wavenumbers larger than the forcing wavenumber $k_f$ and $k^{-5/3}$
scaling for wavenumbers smaller than $k_f$.  Finally \cite{Charney71}
used Kraichnan's ideas to show that similar constraints are placed on
the quasi-geostrophic system due to the joint conservation of
potential enstrophy and energy. Charney showed that in the
quasi-geostrophic system, energy must flow upscale and potential
enstrophy downscale, with predictable scaling exponents of the energy
spectrum in the two ranges. 

The general principle then, is that for a given (inviscidly) conserved
quantity in a system there is a range of scales over which the
flux of that quantity is a constant. The larger the system and the
more separation there is between the source (forcing) and the sink
(dissipation, boundaries), the more extended this universal 'inertial'
range of scales becomes. Once this is established, it is fairly
standard procedure to deduce phenomenologically what the distribution
of such quantities must be in spectral space.

In the inviscid and unforced (${\cal F} = \nu=\kappa =0$, ),
Boussinesq system the potential vorticity $q$ is a local (Lagrangian)
invariant, as is its square the potential enstrophy $Q$, while the
total energy is a global invariant. That is,  
\begin{eqnarray}
\mbox{potential vorticity}~q = \Big(\bm{\omega}_a \cdot \nabla
\rho_T\Big),~~
 \frac{D q}{D t}&=& 0, \nonumber \\
  \mbox{potential enstrophy}~ Q =\frac{1}{2}q^2,~~
\frac{D Q}{D t}= \frac{D}{D t}\int Q~d\bm{x}&=& 0,\nonumber\\
  \mbox{total energy}~E_T = E + P,~~
\frac{ D}{D  t}\int E_T~d\bm{x} &=& 0.\nonumber
\end{eqnarray}
At any point $\bm{x}$, $E=\frac{1}{2}|\bm{u}|^2$ is the kinetic energy,
$P=\frac{1}{2}\theta^2$ is the potential energy of the density
fluctuations. 
The
absolute vorticity $\bm{\omega}_a = \bm{\omega} + f\hat{\bm{z}}$ and
the relative (or local) vorticity $\bm{\omega} = \nabla \times
\bm{u}$. PV may be written in terms of $\theta$ as
\begin{equation}
 q = fN + \bm{\omega}\cdot\nabla\theta + f  \frac{\partial \theta}{\partial z} - N\omega_3.
\label{pv_theta}
\end{equation}
The constant part $fN$ does not participate in the dynamics and we
will therefore neglect it from now on. In what follows we will assume
that $\nu \rightarrow 0$ and $\kappa \rightarrow 0$ such that Prandtl
number $Pr = \nu/\kappa = 1$, and the force $\cal{F}$ is confined to
the lowest modes. Thus we assume a conventional `inertial-range' of
turbulent scales wherein the transfer of conserved quantities
dominates over both their dissipation and forcing. Note that the
global potential vorticity is zero for unbounded flows. While the
potential enstrophy is also Lagrangian invariant, we will use its
global conservation properties in our analysis.

This set of invariants is analogous to those for 2-d turbulence where
the vorticity is Lagrangian invariant and energy and enstrophy are
globally conserved. It therefore seems reasonable to expect that an
analysis similar to that of Kraichnan for 2D turbulence would yield
similarly useful results. The raw Boussinesq system does not permit an
exact relationship between energy and potential enstrophy as in the
case of 2d turbulence. One notices that the potential enstrophy is, in
general, a quartic quantity, while the energy is quadratic. However,
when $f$ and $N$ are sufficiently large, PV linearizes in the
dynamical variables and reduces in fourier space to:
\begin{eqnarray}
\tilde{q}(\bm{k})  \simeq -i (f k_z \tilde{\theta} + N \bm{k}_h \times
  \tilde{\bm{u}}_h) = -i (f k_z \tilde{\theta} + i N k_h \tilde{u}_h)
  \label{qg_pvf}
\end{eqnarray}
where $\tilde{\cdot}$ denotes fourier coefficients, the total
wavevector $\bm{k}$ = $\bm{k}_h + k_z \hat{\bm{z}}$, the horizontal
wavevector component has length $k_h = ({k_x^2 + k_y^2})^{1/2}$, the
vertical wavevector component has length $k_z$ and $\bm{u}_h$ is the
horizontal velocity vector with magnitude $u_h = (u_x^2 +
u_y^2)^{1/2}$. We assume that the vertical velocity $w = u_z \sim 0$
in the leading order, consistent with empirical observations, and use
incompressibility to obtain the last equality of Eq.~(\ref{qg_pvf}).
We will next show how in the formal limits where PV is linear,
asymptotically exact relationships can be extracted between potential
enstrophy and components of the total energy. 

First we briefly review the theoretical results that have been
obtained in the limits in which PV linearizes. For strong rotation and
stratification, \cite{EM98} and \cite{BMN00} showed that the limit as
$Ro \sim Fr = \epsilon \rightarrow 0$ is QG after the inertia gravity
waves are averaged out. The scalings predicted in \cite{Charney71} are
recovered, namely $k^{-3}$ scaling of the energy for $k_f \ll k \ll
k_d$, and $k^{-5/3}$ scaling for $k < k_f$ where $k_f$ is the forcing
wavenumber and $k_d$ is the typical dissipation wavenumber.
\cite{EM98} also showed that the limiting dynamics for $Fr \rightarrow 0$ with
{\it finite} $Ro$ includes the so- called vertically sheared
horizontal flows (VSHF). In both these cases, one can obtain a set of
closed reduced equations for the linear PV dynamics from which all
other quantities of interest can be derived.  There is as yet no
equivalent rigorous theoretical result for the unit-aspect ratio
strongly rotating $Ro \rightarrow 0$ case with finite $Fr$, since in
that limit the scalar equation \ref{bousstta}, which is independent of
$Ro$, vanishes in the leading order. This is a key reason why a
(closed) reduced QG-like set of equations for this limit in the unit
aspect-ratio domain has not been derived. \cite{JulienKMW06} showed
that if an additional small parameter is introduced, namely the
aspect-ratio of the domain, then reduced equations are possible for
the vanishing $Ro$ limit as well.

In recent theoretical work we predicted that the
potential enstrophy associated with the Boussinesq rotating system
exhibits universal statistical properties in the so-called `inertial
range' of scales (\cite{KurienS:Ontcp}) in physical space for
the same limiting regimes of the Rossby and Froude numbers described
above.  Formally, keeping lowest-order terms in a perturbation
expansion in powers of $\epsilon \rightarrow 0$, universal statistics
are found when the potential vorticity is linear in the dynamical
variables $\bm{u}$ and $\theta$. In those cases, the associated
von K\'arm\'an-Howarth equation \cite{KH38} for the two-point
correlation of $q$ yields a simple scaling law for the third-order
velocity-potential-vorticity correlation
\begin{equation}
\langle q(\bm{x})q (\bm{x + r})
(u_L(\bm{x}) - u_L(\bm{x+r}))\rangle = -\frac{2}{3}\varepsilon_Q r
\label{23law}
\end{equation}
where $\bm{r}$ is a physical separation scale, subscript $L$ denotes
the component along $\bm{r}$, and $\varepsilon_Q$ is the mean
dissipation rate of potential enstrophy $Q$.  The $2/3$-law
(\ref{23law}) assumes that the two-point statistics of potential
vorticity have a universal small-scale statistically isotropic
component for statistically steady state flows. A related result was
subsequently derived for the strict QG case (beginning with the
evolution equation for $q_{QG}$) \cite{LindborgE:Thisfr} relating the
third-order moments of potential vorticity difference and horizontal
velocity difference across scale $r$ to potential enstrophy flux,
assuming axisymmetry of the small-scales, and restricted to components of the
velocity that are in the horizontal plane (given that vertical
velocity is strictly zero for pure QG). The numerical experiments to verify 
\ref{23law} are currently underway and will be presented in a separate work.

There have been key numerical studies of the theoretical limits
described above of which we here review the most relevant to our
present study. The limit of strong rotation and stratification
including the case $f = N$ was examined by \cite{B95} in a paper which
studied geostrophic adjustment as a process in which
wave-vortical-wave interactions acted as catalyst for highly efficient
downscale transfer of energy downscale. \cite{SW02} demonstrated the
generation of slow large scales in strongly stratified turbulence with
small scale forcing and the dominance of the $k_h = 0$ (VSHF) wave
modes in the low wavenumbers. \cite{WB04} and \cite{LindborgE:Theecs}
explored the effect of vortical forcing in stratified turbulence
without rotation, thus highlighting the difference that 2D vs. 3D
forcing can make in the evolution of the spectra. \cite{SukSmi08}
showed that fairly small departures from $f/N = 1$ can have measurable
effect on the scaling of the spectra for moderately low-resolution
simulations.  \cite{WB06a, WB06b} and demonstrated the transition to
QG from stratified turbulence as $Ro$ is decreased for fixed small
$Fr$.  \cite{RemSukSmi09} proposed a novel non-perturbative model to
add non-QG effects to a flow and demonstrate the importance of
resolving the small scales (waves) in order to obtain even gross
large-scale features such as the energy growth accurately.

We will here study rotating Boussinesq turbulence both heuristically
and using data from high-resolution simulations for $Ro$ and $Fr$
chosen such that the potential vorticity is nearly linear, and
potential enstrophy is nearly quadratic. In our approach we retain the
type of physically sound arguments used by Kolmogorov, Kraichnan and
Charney, namely a focus on conserved quantities (in the inviscid,
non-diffusive system), existence of inertial ranges where the downscale flux of
a conserved quantity (in our case potential enstrophy) governs the
dynamics and consequent spectral scaling of other related quantities
(energy), and the notion of a statistically universal range of small
scales.

\section{Spectral relationships between potential enstrophy and energy}
We begin by assuming that velocity fluctuations and density
fluctuations are about the same order. We can construct spectral
relationships between the quadratic potential enstrophy $Q(k) =
\displaystyle\frac{1}{2}|\tilde{q}(k)|^2$ and either the horizontal
kinetic energy $E_h(k) = \displaystyle\frac{1}{2}|\tilde{u_h}(k)|^2$ or
the potential energy $P(k) =
\displaystyle\frac{1}{2}|\tilde{\theta}(k)|^2$ based on the parameter
$\Gamma = \displaystyle\frac{f k_z}{N k_h}$, which by definition
resembles a local Burger number in spectra space. From Eq.
\ref{qg_pvf}, for $\Gamma \ll 1$,
\begin{eqnarray}
  \mbox{stratification dominates the PV~}& \tilde{q}(k) \simeq i N k_h \tilde{u}_h \label{pv_Nkhuh},\\
  \mbox{and potential enstrophy~}& Q(k) \simeq \displaystyle \frac{1}{2} |N k_h \tilde{u}_h|^2, \\
  \mbox{giving the constraint~} &\displaystyle\lim_{\kappa_h
    \rightarrow \infty} \int_{\kappa_h}^\infty 
Q(k) d k_h \gg N^2 \kappa_h^2 \int_{\kappa_h}^\infty E_h(k) d k_h \label{Q_rel_Eh}.
\end{eqnarray} 
Eq.~(\ref{Q_rel_Eh}) is interpreted to mean that for wavevectors with
large $k_h$, that is the `wide' wavevectors,
the potential enstrophy $Q(k)$ dominates the downscale (horizontal
scales smaller than $1/\kappa_h$) dynamics and consequently suppresses
the transfer of horizontal kinetic energy into those modes. This is
analogous to the relation in 2d turbulence
\begin{equation}
  \lim_{\kappa \rightarrow \infty} \int_{\kappa}^\infty \Omega(k) dk \gg \kappa^2 \int_{\kappa}^\infty E(k) d k
\label{2d_omegaE}
\end{equation}
wherein the enstrophy $\Omega(k)$ suppresses the downscale transfer of
kinetic energy $E(k)$ for sufficiently high wavenumber $\kappa$, and
in turn forces the inverse cascade of energy.

In the opposite limit
$\Gamma \gg 1$ one can show that,
\begin{eqnarray}
\mbox{rotation dominates the PV~}& \tilde{q}(k) \simeq -i f k_z \tilde{\theta} \label{pv_fkztheta},\\
  \mbox{and potential enstrophy~}& Q(k) \simeq \displaystyle
  \frac{1}{2} |f k_z \tilde{\theta}|^2, \\
 \mbox{giving the constraint~}&\displaystyle \lim_{\kappa_z \rightarrow \infty}
 \int_{\kappa_z}^\infty Q(k) d k_z \gg f^2 \kappa_z^2 \int_{\kappa_z}^\infty P(k) d k_z \label{Q_rel_P}
\end{eqnarray}
which is
interpreted to mean that potential enstrophy suppresses the transfer
of potential energy $P(k)$ into the wavevectors with large $k_z$, that
is the `tall' wavemodes. In physical space this may be loosely
interpreted to mean that potential energy transfer into the small vertical scales is vertical scales. 

More generally, for fixed $f/N$ there is a transition regime when
$k_h/k_z \simeq f/N$ if the wavenumber regime is large enough. Thus,
in an idealized infinite wavenumber flow, for fixed $f/N$ there will
be some regime of wavevectors such that $k_h/k_z \gg f/N$ for which
$\Gamma \ll 1$ and horizontal kinetic energy is suppressed in $k_h$
according to Eq. \ref{kh-5}, and a regime of wavevectors such that
$k_h/k_z \ll f/N$ for which $\Gamma \gg 1$ and potential energy is
suppressed according to Eq. \ref{kz-5}. The transition regime is
defined by a cone in wavevector space with angle $\theta$ to the
vertical such that $\tan \theta = f/N$. All wavenumbers inside the
cone would behave according to $\Gamma \gg 1$ and all those outside
would behave according to $\Gamma \ll 1$. In our simulations in finite
sized boxes, we study three extreme cases: $f/N \ll 1$ where
the cone is extremely narrow (small angle) so that most wavevectors
lie outside the cone giving $\Gamma \ll 1$, $f/N \gg 1$ where the cone
is very broad (large angle) so that most wavevectors lie inside the
cone and $\Gamma \gg 1$ and finally $f/N = 1$ where the cone-angle is
45$\deg$ to the vertical and both $\Gamma \ll 1$ and $\Gamma \gg 1$
regimes exist.

There are several noteworthy differences between the relations
Eq.~(\ref{Q_rel_Eh}) and (\ref{Q_rel_P}) derived above and the 2d
relation Eq.~(\ref{2d_omegaE}). The latter relates two (inviscidly)
conserved quantities in spectral space, while the former relates the
spectrum of a conserved quantity ($Q(k)$) with that of one ($E_h(k)$)
that is not conserved in general. The 2d turbulence constraint forces
enstrophy downscale and energy upscale. In rotating Boussinesq, for
$\Gamma \ll 1$, when the horizontal kinetic energy is suppressed in
the large $k_h$, there is no corresponding constraint in $k_z$ for
either the kinetic or potential energies; conversely for $\Gamma \gg
1$ which potential energy is suppressed in $k_z$, there is no
corresponding constraint in $k_h$.  Therefore there is always the
possibility of a downscale transfer of energy in the unconstrained
wavevector component.  That is, a dual cascade of potential enstrophy
and energy is possible downscale. In addition, Eq.~(\ref{2d_omegaE})
is isotropic in wavenumber in that it depends on the total spherical
wavenumber while our new relations Eqs.~(\ref{Q_rel_Eh}) and
(\ref{Q_rel_P}) are highly anisotropic in wavenumber. As we will show
from the data, the spectra indeed indicate that energy is transferred
downscale in all cases albeit in a highly anisotropic manner.

\subsection{A scaling phenomenology}
Following Kraichnan and Charney one can naively postulate a scaling
exponent for the horizontal kinetic energy and potential energy.
Assuming the potential enstrophy flux governs the downscale dynamics, the
energy scaling in the high wavenumber regime must depend only on the
potential enstrophy dissipation rate $\varepsilon_Q$ and the
wavevector (component) magnitude. Dimensional analysis then results,
for the two limits:
\begin{eqnarray}
  \mbox{$\Gamma \ll 1$:~} E_h(k) &=& C_E\varepsilon_Q^{2/5} 
  k_h^{-3}\\ 
  \mbox{$\Gamma \gg 1$:~} P(k) &=& C_P\varepsilon_Q^{2/5} 
  k_z^{-3}.
\label{k-3} 
\end{eqnarray}
where $C_{E[P]}$ is a prefactor which in principle should depend on
the ratio of rotation to stratification since $E_h$ and $P$ are not
separately conserved. In this study we focus on the scaling in
wavenumber and leave the study of the prefactor to a future work.
Analysis of high-resolution data (see numerical results below and
prior work in \cite{KurWinTay08}) showed that the actual scaling
exponent was much steeper than $-3$, prompting a refinement of the
above scaling estimates as follows.

We first suppose that part of the downscale energy transfer in both
limits is not entirely suppressed. In the case $\Gamma \ll 1$ where we
predict that $E_h(k)$ suppressed for large $k_h$, it may be that
$E_h(k)$ or $P(k)$ is free to transfer into modes with large $k_z$.
And in the case $\Gamma \gg 1$ with $P(k)$ suppressed for large $k_z$,
it may be that $P(k)$ or $E_h(k)$ is free to fill modes with large
$k_h$. Thus there may be a partial flux of the total energy (kinetic
plus potential) to high horizontal or vertical wavenumber leading to
the postulate that the spectral distribution of energy must depend on
both $\varepsilon_Q$ and energy dissipation $\varepsilon_d$ in the
small scales, where the subscript $d$ denotes downscale energy. We
will show {\it a posteriori} that our data supports such a hypothesis.
A simple way to include the dependence on energy dissipation is to
postulate that the energy scaling depends on the {\it ratio}
$\varepsilon_Q/\varepsilon_d$ and on the wavevector component.  Then
dimensional analysis yields:
\begin{eqnarray}
  \mbox{$\Gamma \ll 1$:~} E_h(k) &\propto& C_E\frac{\varepsilon_Q}{\varepsilon_d} k_h^{-5} \label{kh-5}\\ 
  \mbox{$\Gamma \gg 1$:~} P(k) &\propto& C_P\frac{\varepsilon_Q}{\varepsilon_d} k_z^{-5}. 
\label{kz-5}
\end{eqnarray}
Irrespectively of the exact value of the scaling exponent for the steep decay, 
for $\Gamma \ll 1$ the spectral horizontal
kinetic energy distribution becomes independent of $k_z$, while for
$\Gamma \gg 1$ the spectral potential energy distribution becomes
independent of $k_h$. Notice also that this estimate requires that
$\epsilon_Q/\epsilon$ remains finite in the inviscid $\nu \rightarrow
0$, non-diffusive $\kappa \rightarrow 0$ limit. One could presumably
add corrections to mollify the blow-up for small $k_h$ and $k_z$
respectively along the lines of the log-correction derived by
\cite{Kraichnan71b}. This issue of scaling correction 
will not be addressed in this paper
except to say that steeper than -3 scaling implies even stronger
non-local transfers than 2d turbulence as discussed by
\cite{Kraichnan71b}). As we will show in the sections below, our
modified assumptions about energy transfer downscale are supported by
the results of our simulations, and the scaling exponents measured
from our data for the appropriate regimes are remarkably close to
$-5$.

\cite{BGSC06} studies mode-angle dependence of the total energy
spectrum of rapidly rotating turbulence flow, using closure models for
the anisotropic rotating equations for wave turbulence. The flows they
study do not have stratification or any advected scalar and thus have
no potential vorticity. Therefore, our present analysis, which does in
effect study spectra as function of mode-angle via $k_z/k_h$, cannot
be carried over to deduce similar constraints. What we will present
poses several departures from \cite{BGSC06}. First, we address, in
principle, a very broad parameter space of finite $Ro$ and $Fr$ at
high Reynolds numbers via the parameter $\Gamma$. We also, for the
first time, show that potential enstrophy conservation imposes
constraints on the energy distribution in parameter regimes other than
strict QG. And finally we propose a way to understand how a conserved
quantity (potential enstrophy) may affect the spectral distribution of
a non-conserved quantity (horizontal kinetic energy or potential
energy) and furthermore propose a consistent argument to estimate the
resulting spectral scaling exponents.

\section{Numerical simulations and results}
We perform pseudo-spectral calculations of the Boussinesq equations
with rotation on grids of $640^3$ points and $1024^3$ points in unit
aspect-ratio domains. The computational domain has length $L = 1$ to a
side, with wavenumbers in integer multiples of $2\pi$.  The
time-stepping is 4th-order Runge-Kutta with resolution of the smallest
wave frequencies with five timesteps per wave period. In the
simulations at $640^3$, the diffusion of both momentum and density
(scalar) is modeled by hyperviscosity of laplacian to the 8th-power in
order to extend the inertial scaling ranges. The Reynolds number in
these cases is not computed explicitly but the hyperviscosity
coefficient is chosen to resolve the total energy in the largest shell
\cite{Chasnov94,SW02} thus ensuring adequately turbulent flow for a
given resolution. The simulations at $1024^3$ were run with regular
Navier-Stokes viscosity.  The energy input rate $\epsilon_f = 0.5$ and
$1$ respectively for the smaller and larger resolutions, and $Ro$ and
$Fr$ are varied by varying the rotation and stratification rates for
fixed energy input rate which determines the nonlinear timescale. The
forcing is incompressible and equipartitioned between the three
velocity components and $\theta$. This is equivalent to forcing the
two wavemodes and the one vortical modes equally, thus providing an
optimally unbiased calculation. The forcing is also stochastic and
the phases of each mode change sufficiently rapidly that the forced
modes are statistically isotropic. The forcing spectrum is peaked at
$k_f = 4\pm 1$, for large scale forcing. The simulations are dealiased
according to the isotropic two-thirds dealiasing rule.  Lower
resolution runs at $256^3$ and $512^3$ (see \cite{KurWinTay08}) were
also performed in earlier studies, but are not reported here.

We seek to verify our predictions for three particular ``extreme''
regimes in $Ro$, $Fr$ (see Table \ref{param_table}). The first is
strongly stratified with moderate rotation, $f/N \ll 1$ (runs 1 and 4); the second regime has strongly rotating with moderate stratification, $f/N \gg 1$ (runs 2 and 5); and finally the regime which fixes $f = N$ large, (run
3) so that the dependence of potential vorticity on the ratio $\Gamma$
is reduced to a dependence on the cotangent $\tau = k_z/k_h$ alone, of
the angle of the wavevector with respect to the vertical. Presumably, for
sufficiently long inertial ranges any combination of $f$, $N$, $k_h$
and $k_z$ such that $\Gamma \gg$ or $\ll 1$, while maintaining
near-linear PV, may be similarly investigated.  As discussed in
\cite{B95}, strict potential enstrophy conservation is not possible in
a truncated inviscid spectral scheme due to the lack of conservation
of Lagrangian invariants by individual triads. The best we can expect
with this method in the viscous case is that truncation errors are
swamped by viscosity thus giving conserved fluxes in the inertial
range. We mention this fact because we will be using the notion of
potential enstrophy conservation in this study. The spectral method
itself is routinely used to compute flows very similar to those we
study here and and we proceed while remaining aware of the caveats.

For each case in the Table, we will present the calculations to check
specific predictions of the phenomenology proposed, namely the
anisotropic suppression of horizontal kinetic energy and potential
energy in the different $\Gamma$ regimes, and the scaling exponents
predicted in these regimes by Eqs.~(\ref{kh-5}) and (\ref{kz-5}). The spectra 
as a function of $(k_h,k_z)$ are defined as follows:
\begin{eqnarray}
\frac{1}{2}|a(k)|^2 = \frac{1}{2}|a(k_h,k_z)|^2 = \displaystyle\frac{1}{2}\sum_{k_h - 0.5 \leq \kappa_h < k_h + 0.5}| a(\kappa_h, k_z)|^2
\end{eqnarray}
where $a$ is any of $\tilde q$, $\tilde{u}_h$ or $\tilde{\theta}$.
We will present the horizontal kinetic energy and
potential energy spectra (a) as a function of $k_h$ for various values
of fixed $k_z$, (b) as a function of $k_h$, summed over $k_z$, (c) as
a function of $k_z$ for various values of fixed $k_h$, and (d) as a
function of $k_z$, summed over $k_h$. The (a)-(d) above correspond to
subplots (a)-(d) in the sets of figures to follow. In each subplot,
horizontal kinetic energy is plotted on top and potential energy on
the bottom. We will also show the more conventional decomposition of
the energy in contributions from the linear eigenmodes of the
Boussinesq system. There are two inertial-gravity
wave-eigenmodes with non-zero frequency and zero PV which contribute
to the wave (or ageostrophic) energy spectrum. The remaining eigenmode
has zero frequency and linear PV and contributes to the vortical (or
geostrophic) energy spectrum. We will attempt to relate the mode-angle
representation to the usual wave-vortical decomposition.  Since this
paper is largely an exploration of spectral scaling in the large
wavenumbers, we will, for clarity's sake, use the hyperviscous
simulations data with their extended ranges to compute scaling
exponents. The Navier-Stokes cases (runs 4,5) show no appreciable
differences from their hyperviscous counterparts except for shortened
scaling ranges, as is to be expected. Wherever appropriate, and when
it does not result in a proliferation of figures, we will also present
the data from the simulations with Navier-Stokes viscosity.

\begin{table}
\centering
\begin{tabular}{cccccll}
run & grid  & diffusion &$\epsilon_f$  & $k_f$ &$f$ ($Ro$) & $N$ ($Fr$) \\
1 & $640^3$ & hyper&0.5 & 4 & 14 (0.48) & 3000 (0.0023) \\
2& $640^3$ & hyper&0.5 & 4 & 3000 (0.0023) & 14 (0.048) \\
3  & $640^3$ & hyper&0.5 & 4 & 3000 (0.0023) & 3000 (0.0023) \\
4 & $1024^3$ & NS & 1 & 4& 8.58 (1) & 8580 (0.001) \\
5 & $1024^3$ & NS & 1 & 4&4290 (0.002) & 8.58 (1) \\
\end{tabular}
\caption{The parameters for the numerical simulations of the 
  Boussinesq equations. The notation ``hyper'' denotes hyperviscous
  diffusion of Laplacian to the eighth power, applied to both momentum and scalar; ``NS'' denotes regular Navier-Stokes viscosity.}
\label{param_table}
\end{table}

\begin{figure}
\centering
\includegraphics[scale = 0.35]{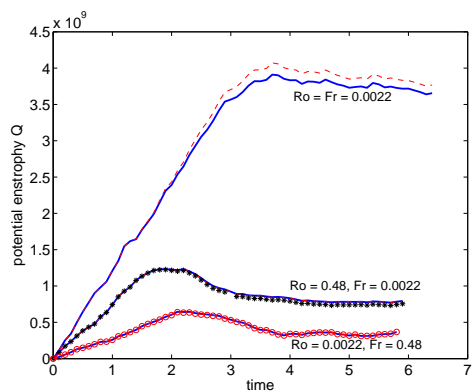}
\caption{[Color online] Global potential enstrophy evolution in time for runs 1,2, and 3. 
The total $Q$ and its quadratic components corresponding to linear $q$, are 
shown. Solid line: total $Q$; dashed line : $\frac{1}{2}|f {\partial \theta
  \over \partial z} + N \omega_3|^2$; stars: $\frac{1}{2}{|N \omega_3|^2}$; 
circles: $\frac{1}{2}{|f {\partial \theta \over \partial z}|^2}$. In
each limiting case, the corresponding quadratic part of the potential
enstrophy captures the total Q to within 3\%.
\label{potens}}
\end{figure}
First we show in Fig. \ref{potens} that in each of the three cases,
the global potential enstrophy (solid line) is captured by one half of
the square of the corresponding linear PV, to within 3\% or better
agreement.
In each case, the global
potential enstrophy, after a period of spin-up and over-shoot, settles
to a constant, indicating statistically steady state for this
quantity, the mean rate of input of potential enstrophy balancing the
mean rate of dissipation. We are thus able to stablish the time-frame
over which the small scales, governed by potential enstrophy dynamics,
have achieved statistically steady state. The small scale spectra of
energy that we are interested in, achieve a statistically steady
state. The total energy meanwhile (not shown) continues to grow since
it can transfer upscale, where in our case, there is no sink to drain
the energy. The calculation of all statistical quantities in what
follows occurs over the time-frame where the mean potential enstrophy
is constant, $3.5 < t < 6$, over about 25 frames.

\subsection{Strong stratification, moderate rotation}\label{Ro1Fr0}
\begin{figure}
\subfigure[Spectra of horizontal kinetic energy (solid lines) and potential energy (dashed lines) as a function of $k_h$ for various fixed $1 \leq k_h \leq 100$ as indicated.]{
\includegraphics[scale = 0.28]{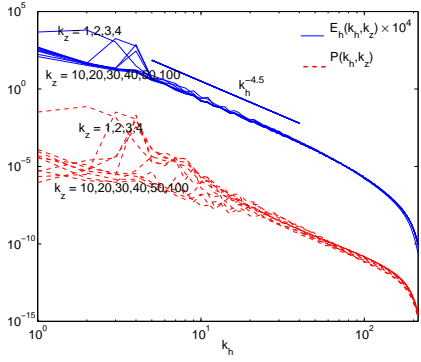}
\label{Ro1Fr0_spec:a}
}
\subfigure[Spectra of horizontal kinetic energy (solid line) and potential energy (dashed line), as functions of $k_h$, after summing over $k_z$.]{
\includegraphics[scale = 0.28]{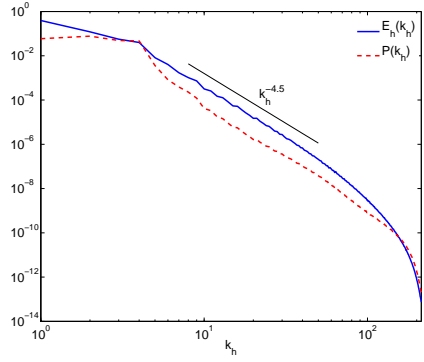}
\label{Ro1Fr0_spec:b}
}
\subfigure[Spectra of horizontal kinetic energy (solid lines) and potential energy (dashed lines) as a function of $k_z$ for various fixed $1 \leq k_h \leq 30$ as indicated.]{
\includegraphics[scale = 0.28]{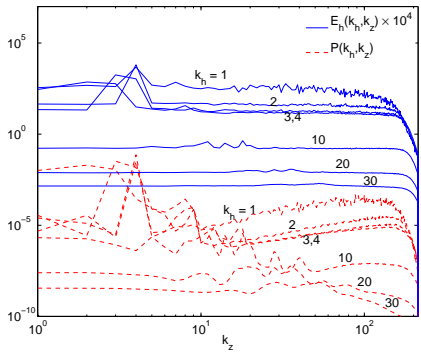}
\label{Ro1Fr0_spec:c}
}
\subfigure[Spectral of horizontal kinetic energy (solid line) and potential energy (dashed line), as functions of $k_h$, after summing over $k_z$.]{
\includegraphics[scale = 0.28]{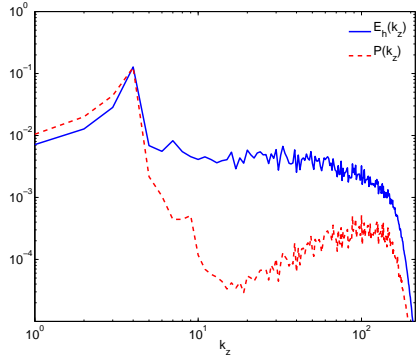}
\label{Ro1Fr0_spec:d}
}
\caption{[Color online] The horizontal kinetic energy and potential energy spectra
  for $f \ll N$ moderately rotating and strongly stratified flow
  (Table \ref{param_table} run 1). \label{Ro1Fr0_spec} }
\end{figure}
This case is studied using data from runs 1 and 4 of Table
\ref{param_table}. We are in the limit $\Gamma \ll 1$ since nearly all
wavevectors lie outside the narrow cone with vertical angle $tan
\theta = f/N = 0.0047$. Therefore, $\tilde{q}(k) = \tilde{q}(k_h, k_z)
\sim N k_h \tilde{u}_h$ irrespective of $k_z/k_h$ and the
spectral scaling of $E_h(k)$ should become independent of $k_z$ as
given by Eq. \ref{kh-5}. Fig.~(\ref{Ro1Fr0_spec:a}) (top) and shows
that $E_h(k_h,k_z)$ collapses to a function of $k_h$ alone over the
wide range $0 < k_z < 100$, and scales very close to $k_h^{-5}$. In
fact, the collapse is so complete that the converse plotting of $E_h$
as a function of $k_z$ in Fig.  (\ref{Ro1Fr0_spec:c}) for various
fixed $k_h$, gives a series of plateaus, indicating no $k_z$
dependence at all in the inertial range for any particular $k_h$.  The
corresponding plots for $P(k_h,k_z)$ shows that it nominally
follows the scaling of $E_h$ as a function of $k_h$ (Figs.
\ref{Ro1Fr0_spec:a}), yielding $k_z$-averaged scaling of $k_h^{-5}$
(Fig. \ref{Ro1Fr0_spec:b})).  However the converse plotting of $P$
(Fig. \ref{Ro1Fr0_spec:c}) shows that it does not become completely
independent of $k_z$ even though, when {\it summed} over $k_z$ (Fig.
\ref{Ro1Fr0_spec:d}), its scaling becomes close to $k_h^{-5}$ as
predicted in this limit for $E_h(k_h)$.  Thus $P$ appears to mimic the
scaling of $E_h$ as a function of $k_h$, but not as a function of
$k_z$. 

We now make a more conventional representation of the spectra by
plotting the energy contained separately in the wave and vortical
modes as a function of the spherical wavenumber $k$. 
\begin{figure}
\centering
\includegraphics[scale=0.3]{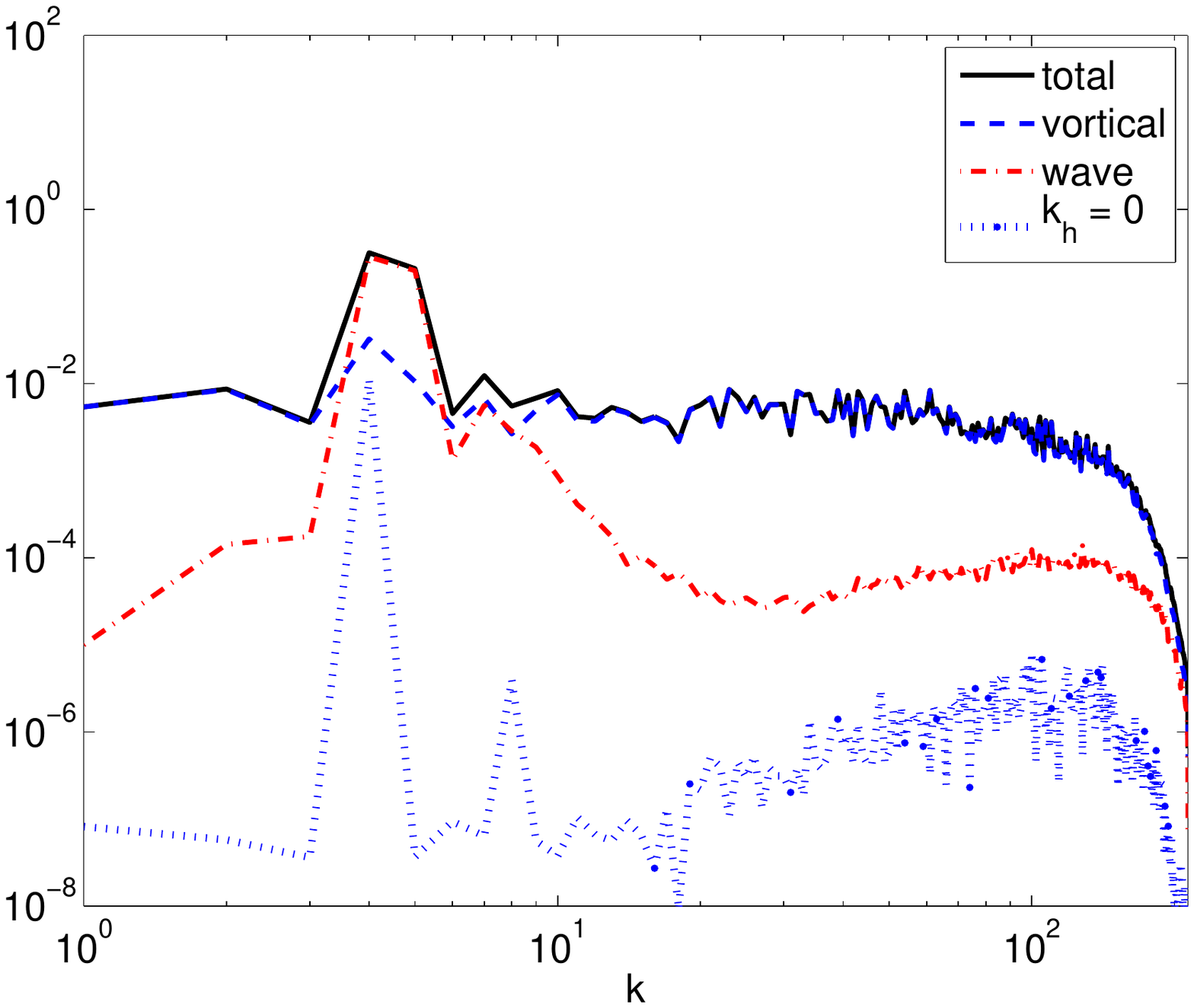}
\includegraphics[scale=0.3]{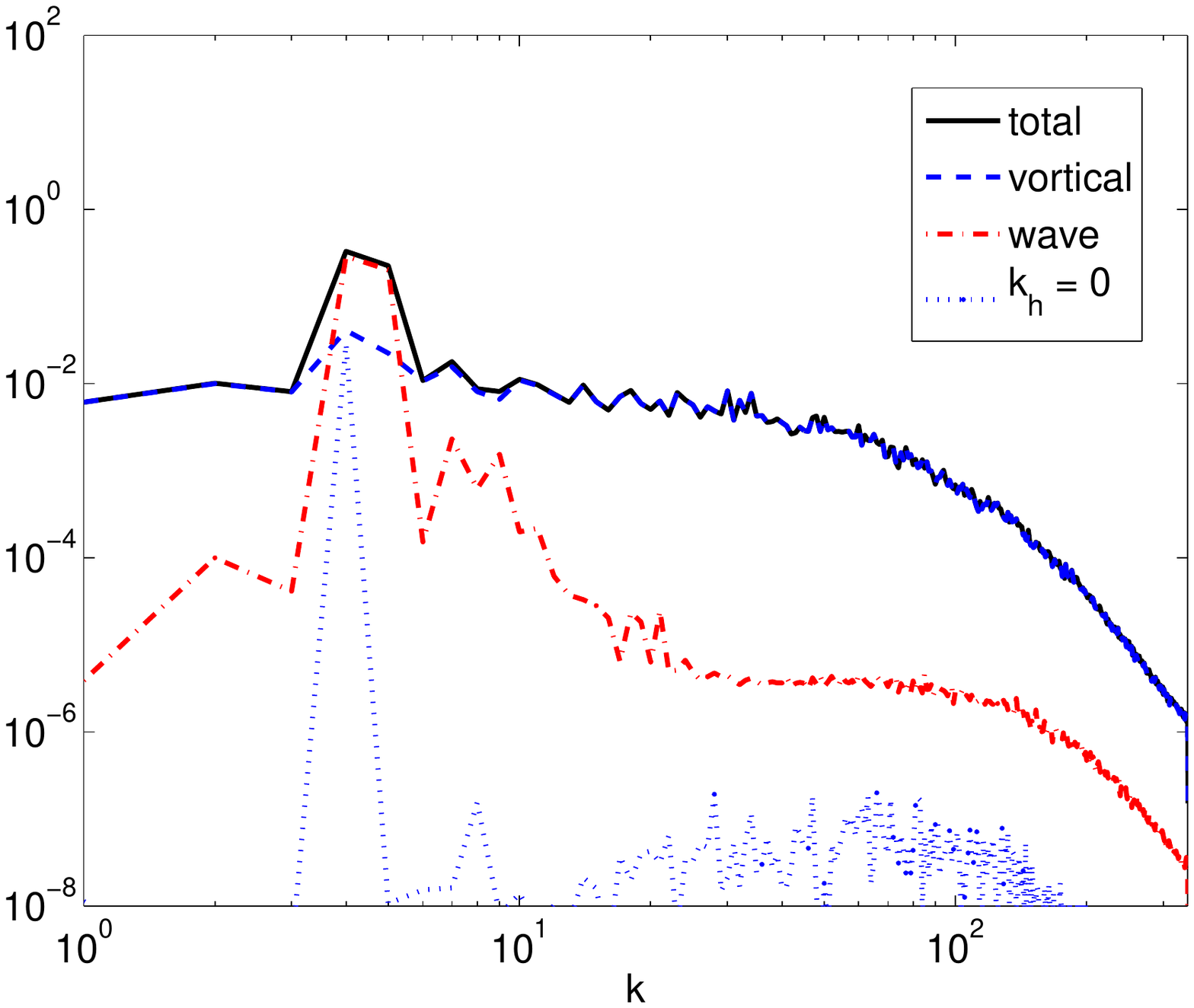}
\caption{[Color online] Strongly stratified and moderately rotating cases, with
  hyperviscosity (run 1, left) and with Navier-Stokes viscosity (run
  4, right). Total (black solid), wave (red dash), vortical (blue
  dot-dash) and $k_h = 0$ mode (blue dotted) energy spectra.  Almost
  the entire total energy is captured by the vortical modes within
  which the energy is uniformly distributed as $k^0$.  The wave-mode
  contribution dominates in the forced modes only. The $k_h = 0$ modes
  form a negligible contribution, highlighting the difference from
  VSHF. \label{Run1_WVT}}
\end{figure}
Figure \ref{Run1_WVT} shows that almost the entire downscale energy is
captured by the vortical modes. The narrow upscale range is also
filled by the vortical modes. The total energy (vortical energy)
distribution is nearly a constant as a function of $k$. The wave
energy is subdominant everywhere except in the forced modes,
indicating that the wave energy does not move efficiently out of those
modes in this strongly stratified regime.  This figure is to be
compared with Fig. \ref{Ro1Fr0_spec:d} to see that for $k > k_f$ the
vortical modes mimic the horizontal kinetic energy and that the
independence of the latter from $k_z$ translates to a constant
spectral distribution over $k$ as well, given that the dependence of
$E_h$ on $k_h$ is rapidly very decaying (Fig.  \ref{Ro1Fr0_spec:b}).
To the best of our knowledge, this is the first observation and
explanation of the asymptotic high wavenumber energy spectrum for
strongly stratified flow with finite rotation, with isotropic forcing
in the low-wavenumbers (large scales).  \cite{WB04} observed
$k_h^{-5}$ scaling for the total energy spectrum and the vortical mode
energy spectrum in flows with rather low stratification $N$, zero
rotation and forced in the vortical modes only. They also observed
saturation in $k_z$ of the total energy spectrum. In the light of the
analysis in the present work, it would appear they achieve something
like the regime $\Gamma \ll 1$ and linear $q \simeq N\omega_3$, not
with large $N$, but with large $\omega_3$ achieved by forcing vortical
modes only. Their paper \cite{WB06b} which studies similar parameter
regimes in $Fr$ while varying $Ro$ does show $k_z^0$ scaling of the
vortical mode energy for $Ro \sim {\cal O}(1)$ but does not recover
the $k_h^{-5}$ steep decay. As we have pointed out above, the exact
decay exponent does not matter as long as it is a decay and is
independent of $k_z$, leading to the $k_z^0$ scaling.  Our analysis
goes a step further in proposing an explanation of {\it why} there is
suppression in $k_h$ and and not in $k_z$ due to the very particular
constraints imposed on energy by the potential enstrophy, leading to
independence from $k_z$. We have also conclusively shown that the
asymptotic small $Fr$ regime for the small scales can be achieved with
unbiased isotropic forcing and in the presence of finite rotation.

There is also a stark difference from the vertically sheared
horizontal flows (VSHF) ($k_h = 0$ modes) expected to be included in
this limit when the forcing is at {\it high} wavenumbers (small
scales). In our data the $k_h = 0$ modes contain a very small
percentage of the total energy (see Fig. \ref{Run1_WVT}) and the vortical (PV)
modes contain most of the energy, whereas the opposite is true for the low
wavenumbers in the case where the forcing is in the high wavenumbers
(\cite{SW02,EM98}).

\subsection{Strong rotation, moderate stratification, $Ro = 0.0022$,
  $Fr = 0.48$}
\begin{figure}
\subfigure[]{
\includegraphics[scale = 0.28]{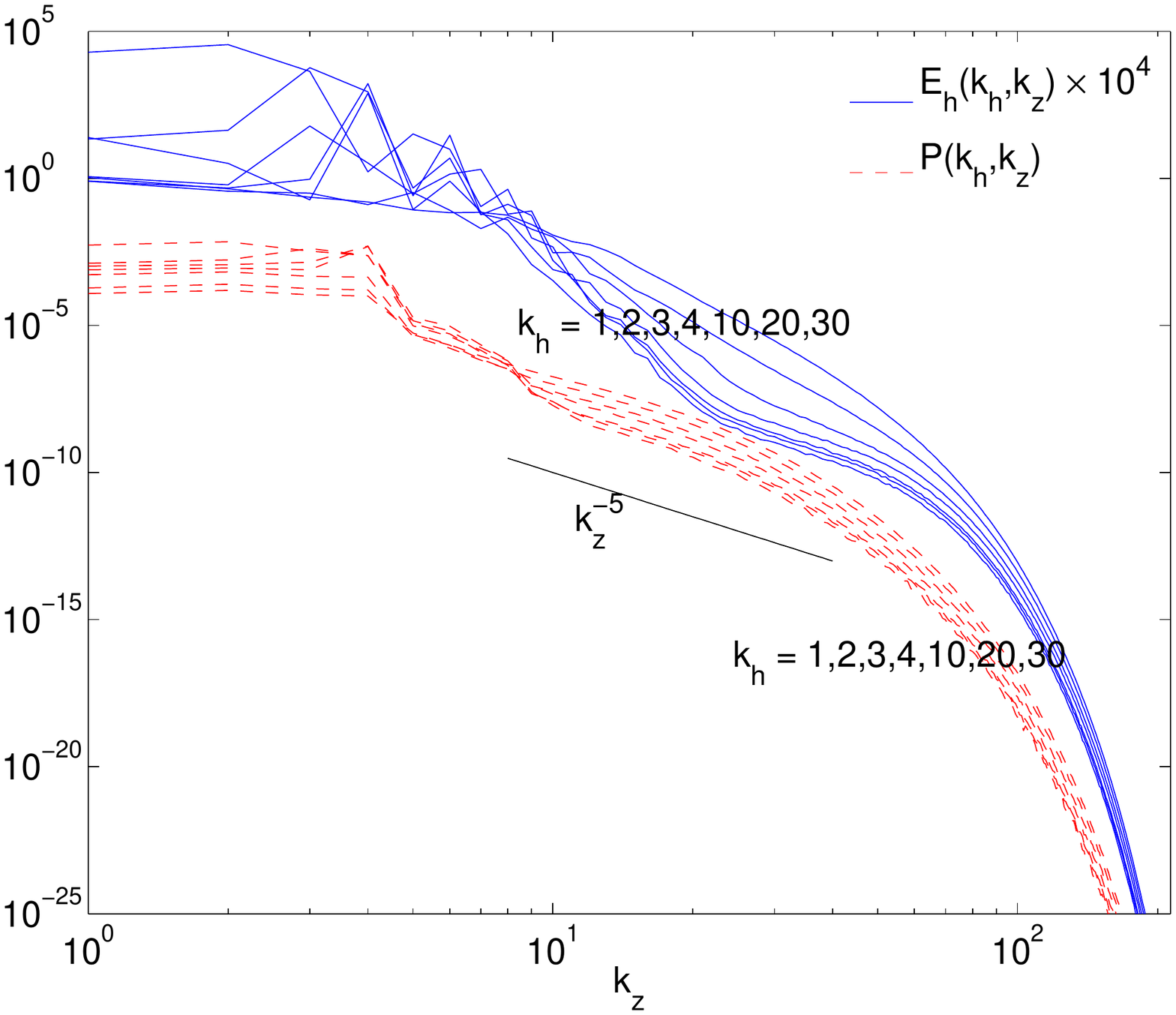}
\label{Ro0Fr1_spec:a}
}
\subfigure[]{
\includegraphics[scale = 0.28]{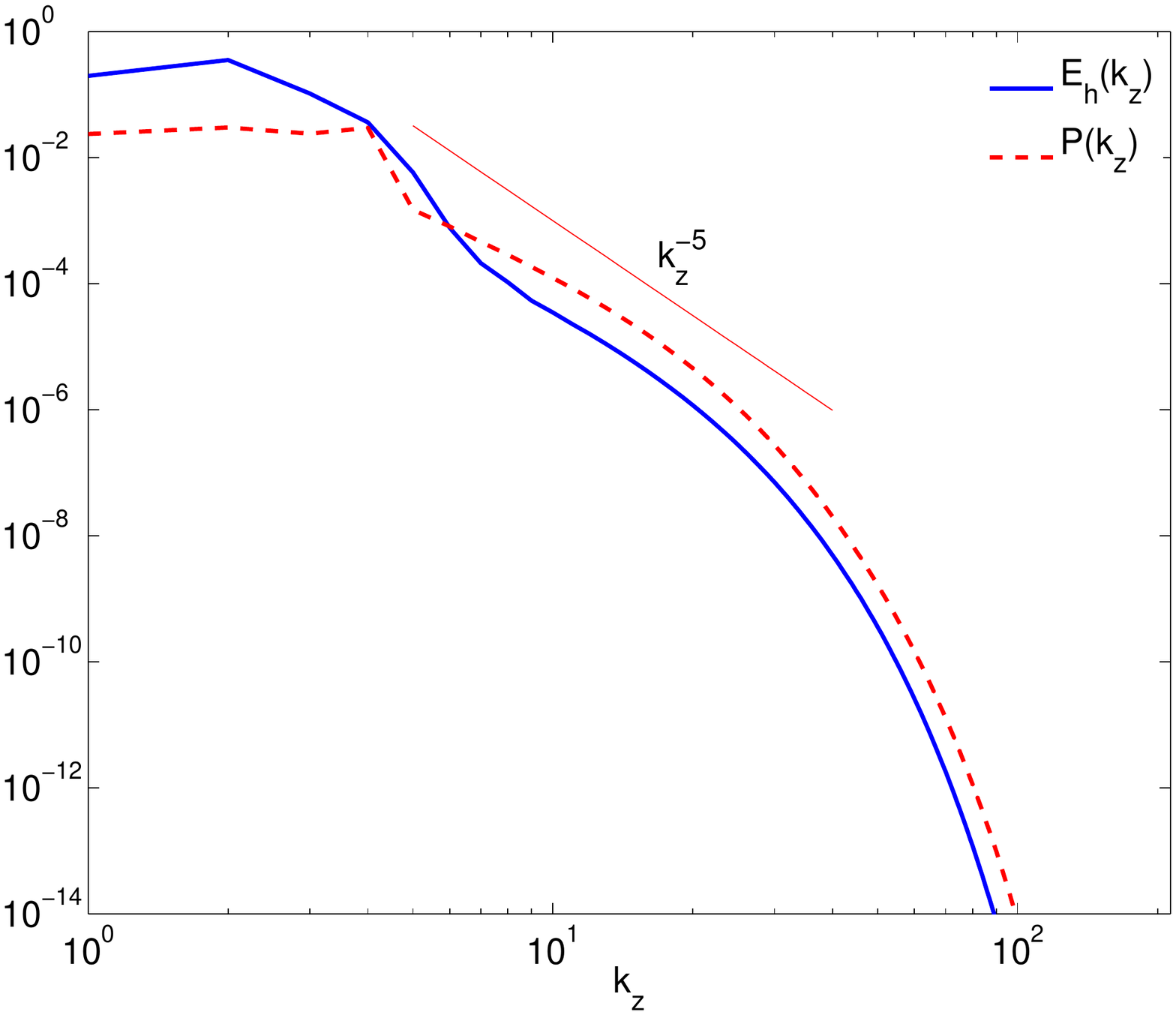}
\label{Ro0Fr1_spec:b}
}
\subfigure[]{
\includegraphics[scale = 0.28]{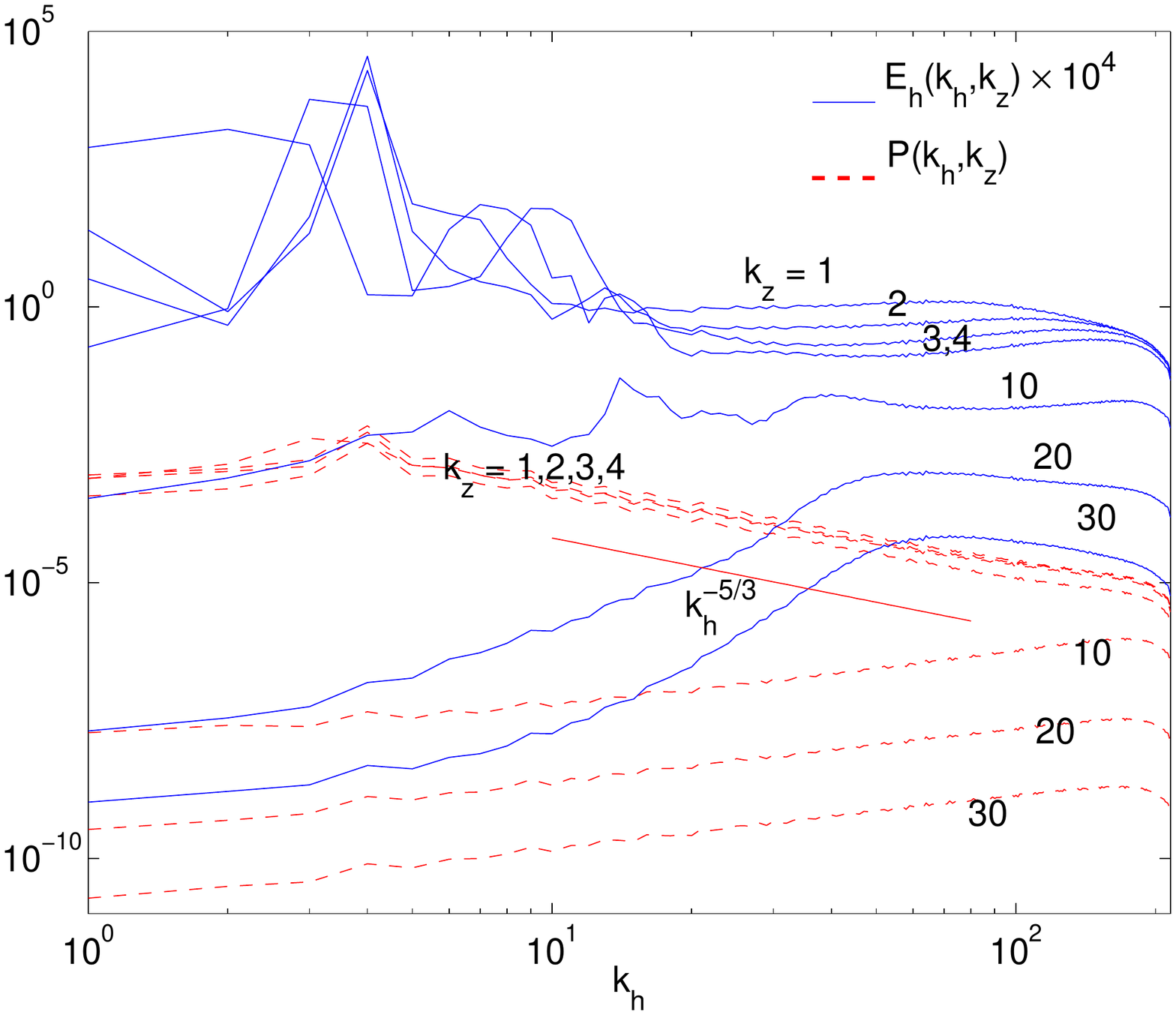}
\label{Ro0Fr1_spec:c}
}
\subfigure[]{
\includegraphics[scale = 0.28]{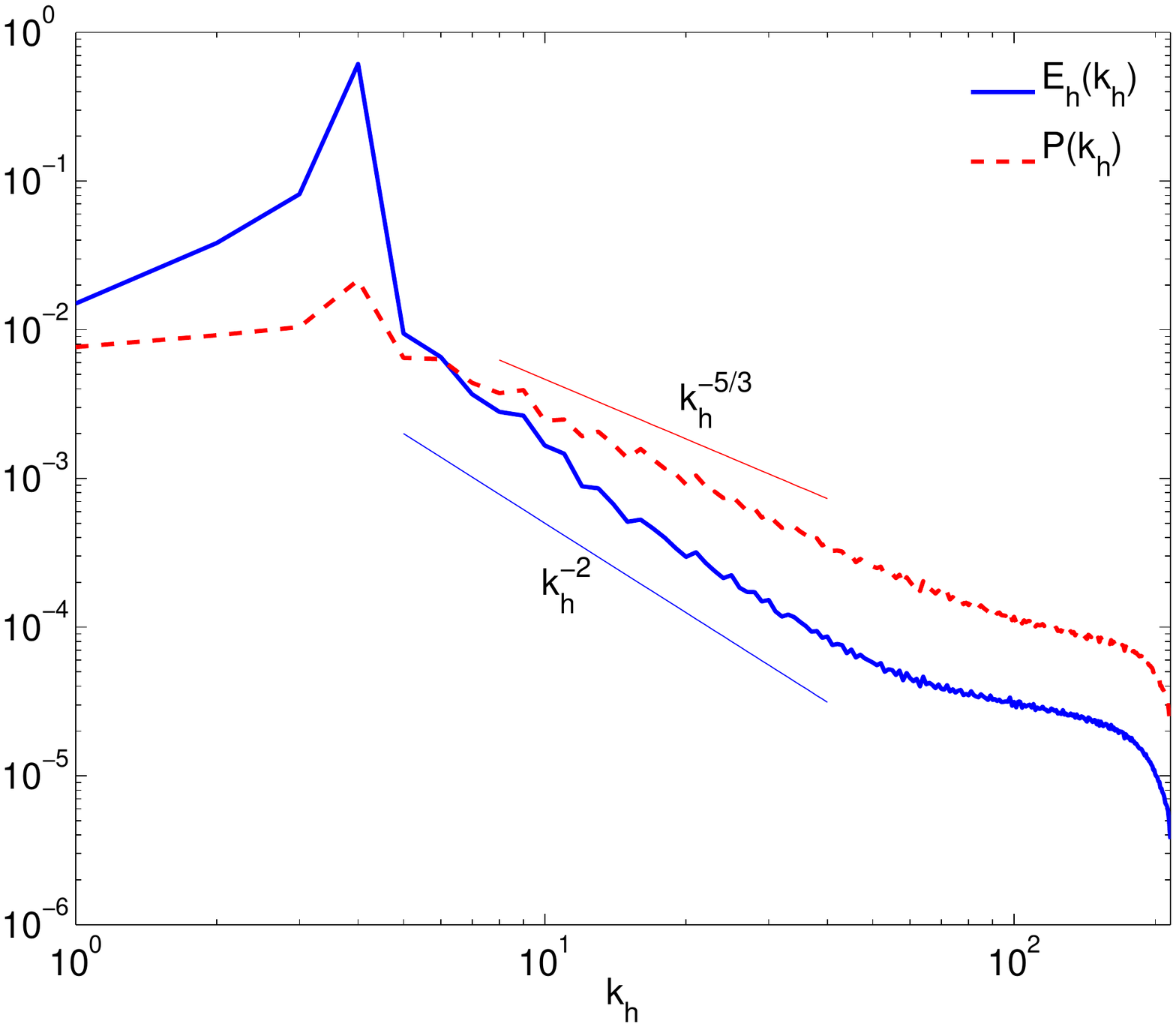}
\label{Ro0Fr1_spec:d}
}
\caption{[Color online] The horizontal kinetic and potential energy spectra for $f \gg N$
  strongly rotating and moderately stratified flow (Table \ref{param_table}
  (ii)). The quantities plotted for subfigures (a,b,c,d) are the same as 
those for Fig. \ref{Ro1Fr0_spec}(c,d,a,b) respectively. \label{Ro0Fr1_spec}}
\end{figure}
For run 2 of our simulations with moderate stratification and strong
rotation, $\Gamma \gg 1$ and we expect to observe Eq. (\ref{kz-5}). In
Fig.  \ref{Ro0Fr1_spec:a} we see that although the scaling of $P(k)$
in $k_z$ for all $k_h$ shows some degree of agreement, the curves
themselves have not collapsed. This is also shown by the converse
plotting of $P(k)$ for various $k_z$ functions of $k_h$ in Fig.
\ref{Ro0Fr1_spec:a}. However we note that the trend as $Ro$ decreases
is for the potential energy curves to collapse to functions of $k_z$
alone. Thus $Ro=0.002$ may not be sufficiently small to see the
limiting small-scale dynamics for $Ro \rightarrow 0$, $Fr \sim {\cal
  O}(1)$.  And this despite the fact that potential enstrophy has
converged to the stratification dependent quadratic component as seen
in Fig.  \ref{potens}. Thus it appears, at least empirically, that for
the small scales, the asymptotic regime of strong stratification is
easier to achieve than that of strong rotation. This would seem
related to the fact that the $Ro \rightarrow 0$ limit for finite $Fr$
is not amenable to a reduced set of equations since in that limit 
the scalar evolution equation vanishes in the leading order.

Figures \ref{Ro0Fr1_spec:b} and \ref{Ro0Fr1_spec:d} show that the
$k_z$-summed spectra scale between $k_h^{-2}$ and $k_h^{-4/3}$ and
the $k_h$-summed quantities scale approximately as $k_z^{-5}$. Thus,
on average both horizontal kinetic and potential energies are
suppressed in the large $k_z$ modes while both are free to fill the
large $k_h$ modes. Again this justifies the ansatz that the scaling
depends on $\varepsilon_Q/\varepsilon_d$, with downscale fluxes of both
potential enstrophy and energy fluxes playing a role in the final
development of the high wavenumber spectra.

\begin{figure}
\centering
\includegraphics[scale=0.3]{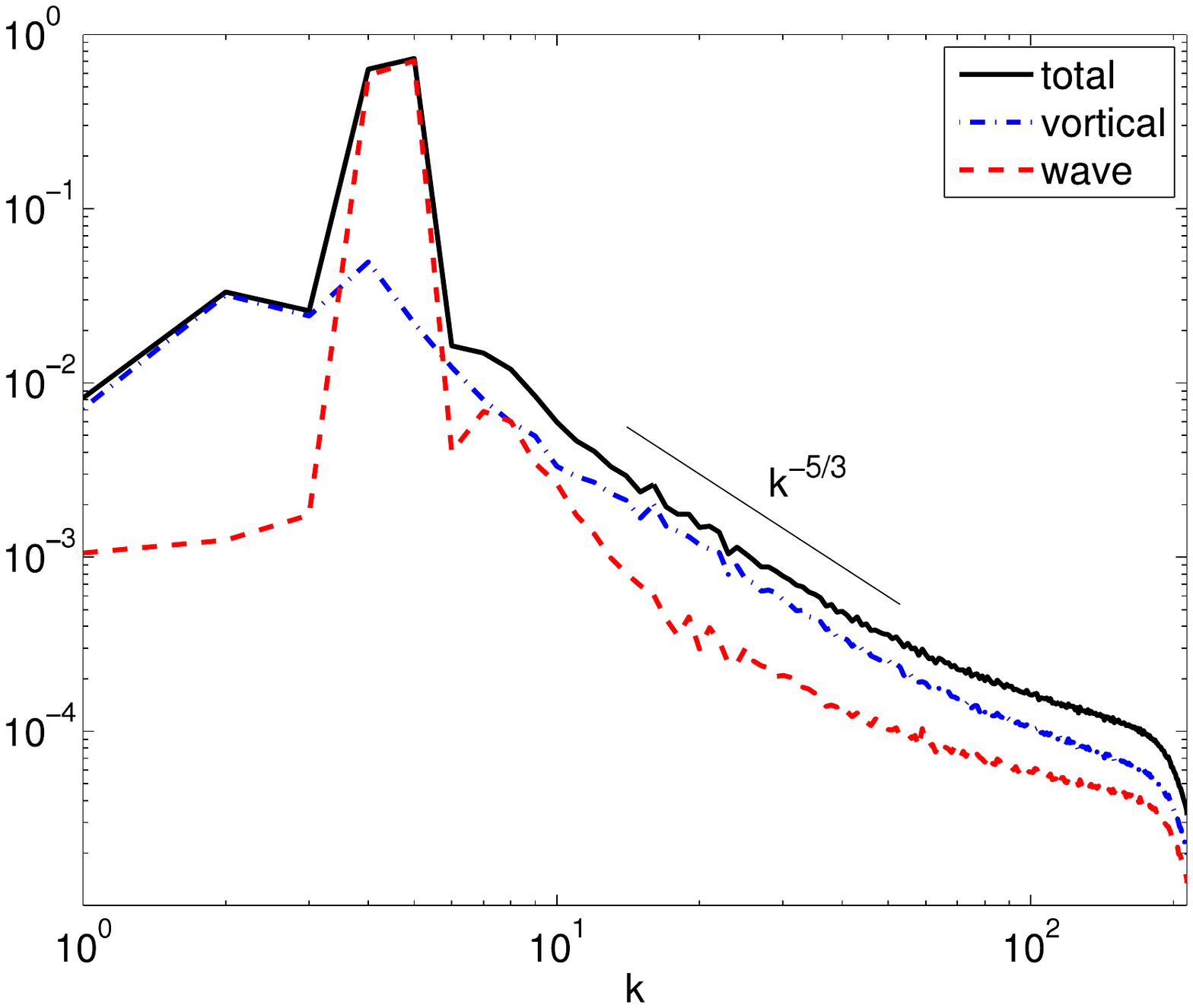}
\includegraphics[scale=0.3]{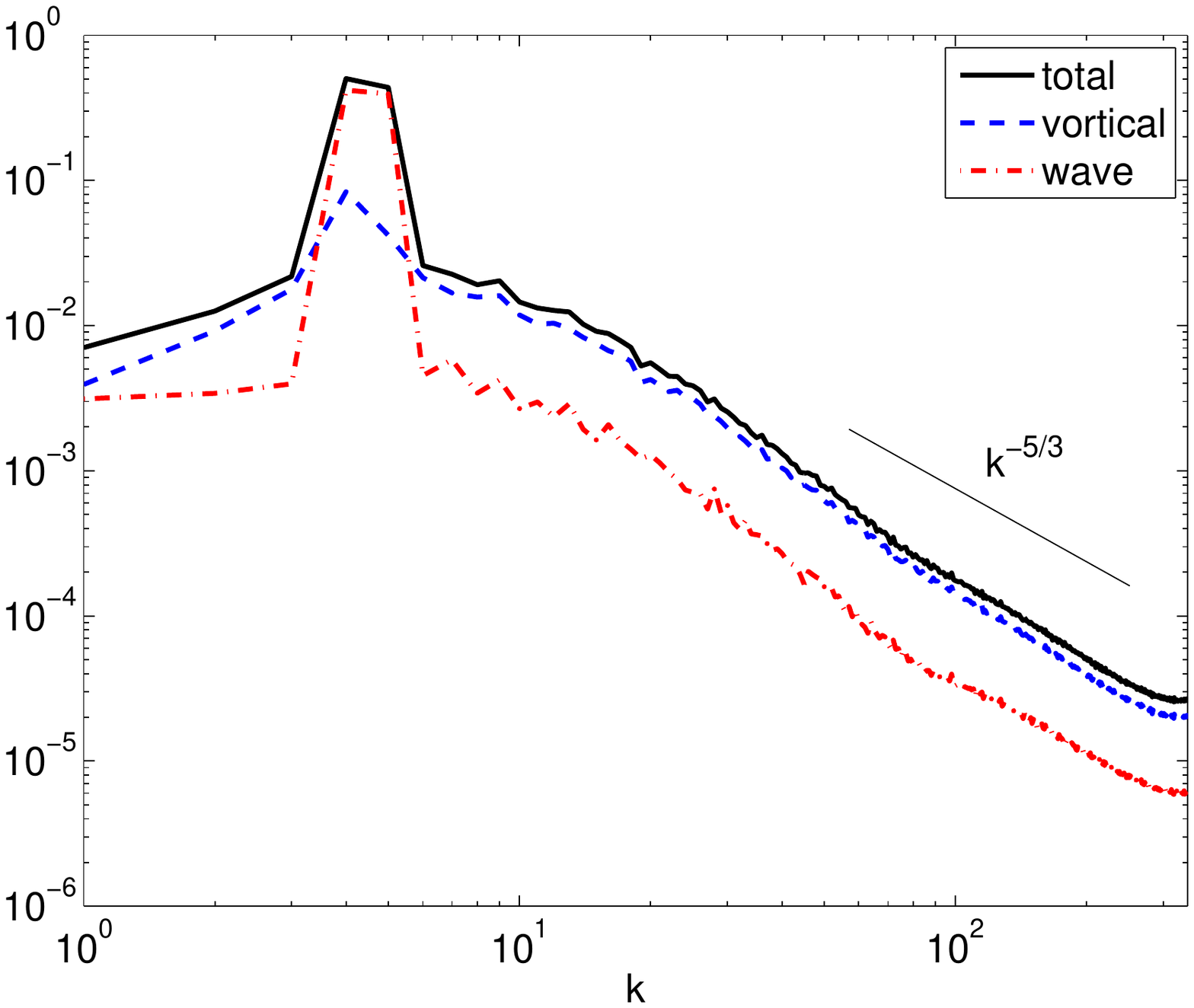}
\caption{[Color online] Strongly rotating and moderately stratified cases, with
  hyperviscosity (run 2, left) and with Navier-Stokes viscosity (run
  5, right). Total (black solid), wave (red dash), vortical (blue
  dot-dash) energy spectra.  Almost the entire total energy is
  captured by the vortical modes within which the energy is uniformly
  distributed approximately as $k^{-5/3}$.  The wave-mode contribution
  dominates in the forced modes only.
\label{Run2_WVT}}
\end{figure}
The conventional representation of wave and vortical mode spectra,
Fig. \ref{Run2_WVT} shows that in this regime, the vortical modes once
again dominate the contribution to the total energy at all wavenumbers
except around the forced wavenumbers, where the wave-mode contribution
dominates.  At high wavenumbers, the vortical modes scale more or less
as $k^{-5/3}$. When compared with Fig. \ref{Ro0Fr1_spec:d} this may be
interpreted as the vortical modes being dominated by the potential
energy, which persists strongly into the large $k_h$ modes while
decaying rapidly in the large $k_z$ modes. Thus in the aggregate the
$k_h$ scaling dominates in the spherically averaged spectra as well.
On the right of Fig. \ref{Run2_WVT} shows the same decomposition for
the run 5 in the strongly rotating regime which shows qualitative
agreement with the hyperviscous case, comparable scalings and hence
providing some confidence that in these cases hyperviscosity does not
bias our interpretation with undesirable artifacts.

The point of similarity between the strongly stratified versus the
strongly rotating limits is that in both cases the vortical modes
dominate the high wavenumbers or small scales. However, in the former
the vortical modes carry the horizontal kinetic energy whereas in the
latter the vortical modes carry the potential energy. This is very
different from QG wherein the vortical mode energy decays as $k^{-3}$ in
the high wavenumbers. Note also the difference from the equally
strongly rotating and stratified case (see below) which, as is well
known (see \cite{B95}), carries most of the high wavenumber energy in
the wave modes which scale as $k^{-1}$.

\subsection{Equally strong rotation and stratification, $Ro = Fr \sim 0.0022$}
In the special case of $f = N$ the dependence of PV on $\Gamma$
reduces to a dependence on the cotangent $\tau = k_z/k_h$ of the
wavevector angle. As $\tau \ll 1$, the $N$-dependent part prevails
while for $\tau \gg 1$, the $f$-dependent part takes over. That is,
the cone in $k$-space separating $\Gamma \ll 1$ and $\Gamma \gg 1$
makes a 45$^o$ angle to the vertical allowing equal numbers of modes
in both regimes.
\begin{figure}
\centering
\includegraphics[scale=.35]{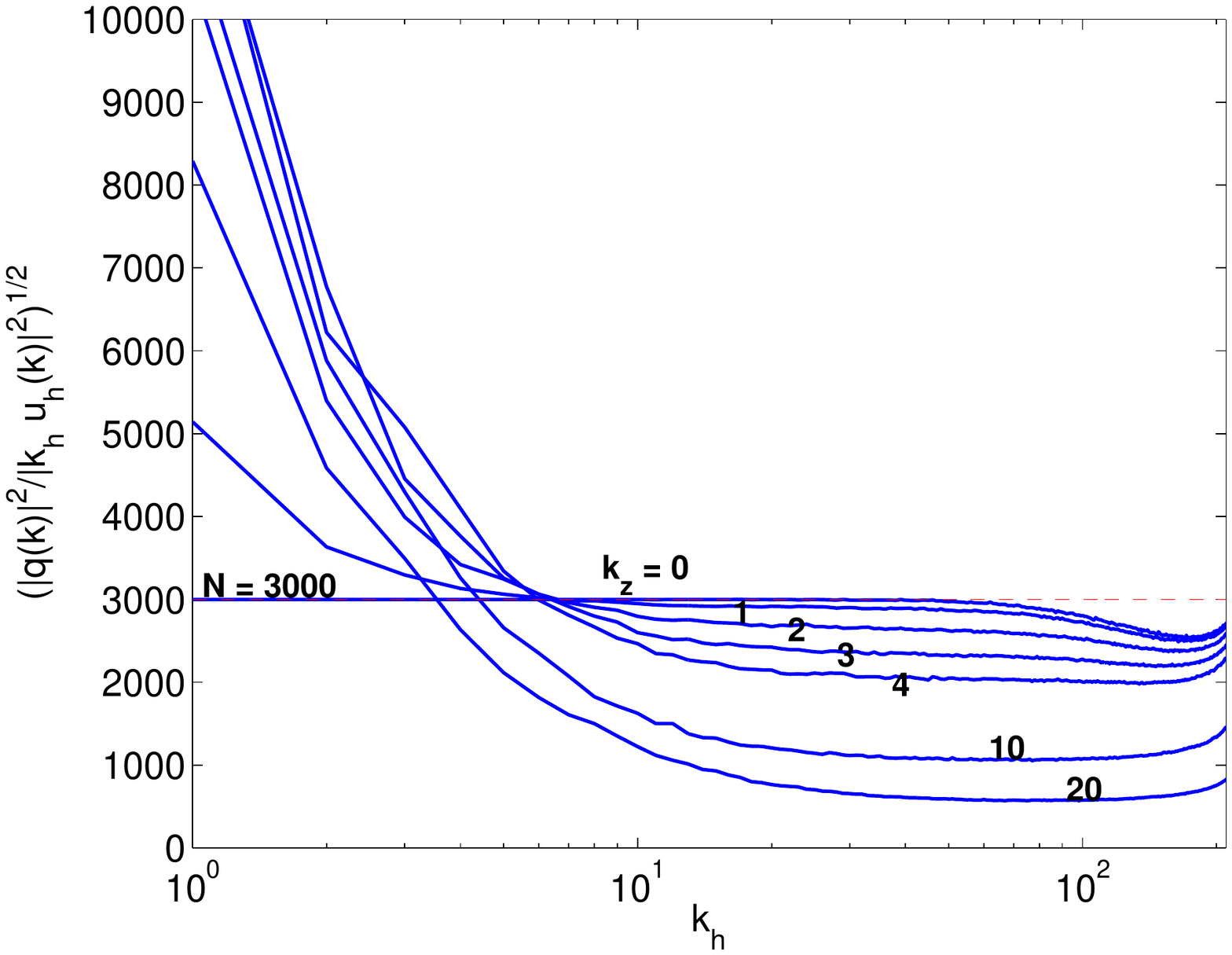}
\includegraphics[scale=.35]{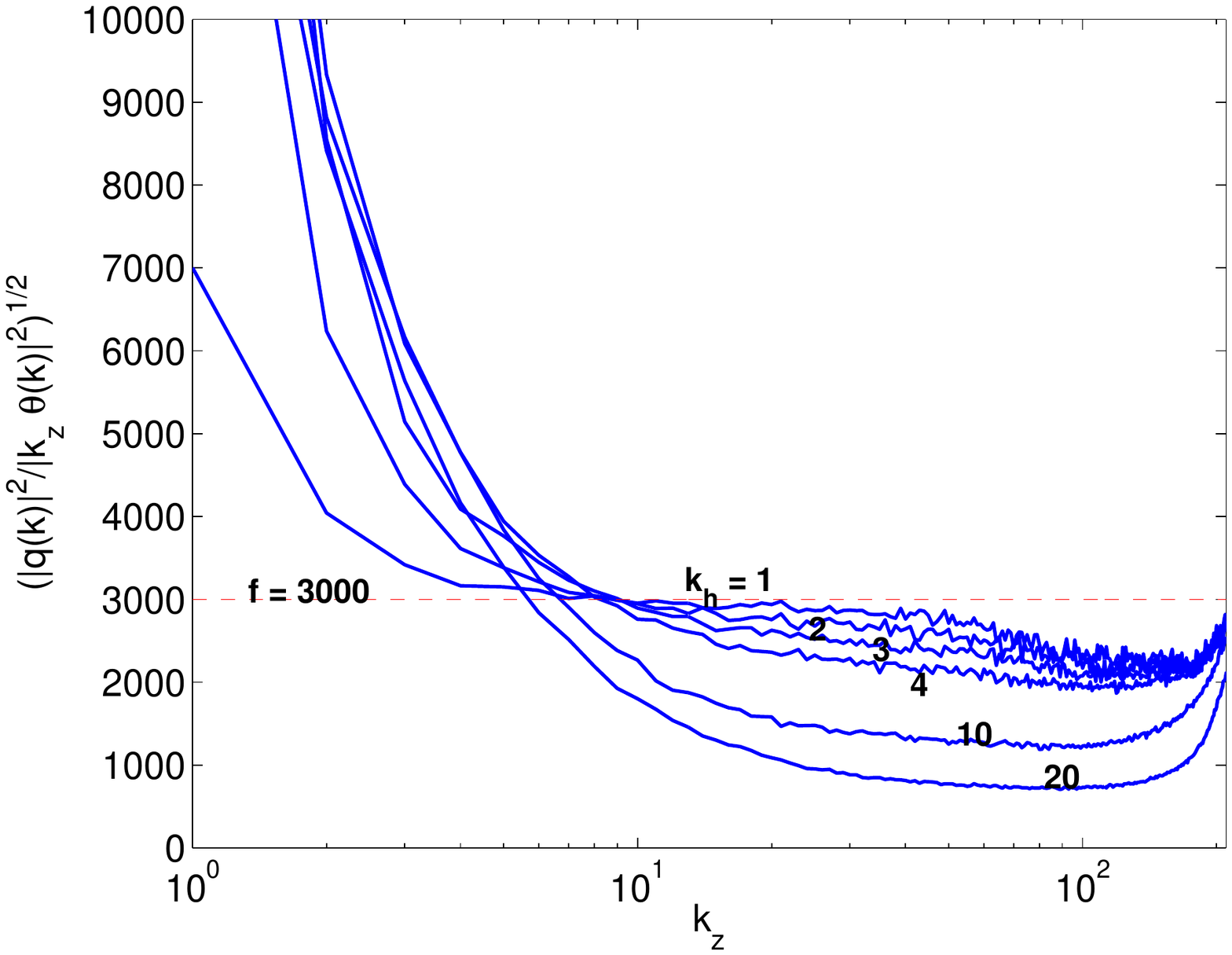}
\caption{[Color online] Run 3, $f = N = 3000$ ($Ro = Fr = 0.0022$).  Left:
  $|\tilde{q}(k)|$ normalized by $|k_h \tilde{u}_h(k)|$ as a function
  of $k_h$ for various values of $k_z$.  The normalized potential
  vorticity asymptotes to $N=3000$ for $\Gamma = \tau \ll 1$ as
  predicted in Eq.  (\ref{pv_Nkhuh}). Right: $|\tilde{q}(k)|$
  normalized by $|k_z\theta(k)|$ as a function of $k_z$ for various
  values of $k_h$.  The normalized potential enstrophy asymptotes to
  $f=3000$ as $\Gamma = \tau \gg 1$ as predicted in Eq.
  (\ref{pv_fkztheta})}
\label{Qnormfigs}
\end{figure}
Using run 3, we show that as $\tau \ll 1$ the value of $|\tilde{q}(k)|
\simeq N |k_h \tilde{u_h}(k)|$ (Fig. \ref{Qnormfigs}, left panel),
while as $\tau \gg 1$ the value of $|\tilde{q}(k)| \simeq f |k_z
\tilde{\theta}(k)|$ (Fig. \ref{Qnormfigs}, right panel).  Since there
is thus an average over the annulus of horizontal wavenumbers
contained in $k_h \pm 0.5$ the curves on the left panel of Fig.
\ref{Qnormfigs} are smoother than those on the right. The special case
of $k_h = 0$ has been omitted for clarity on the right panel as it is
noisy (less averaging), while the $k_z = 0$ case on the left poses no
such problems. This figure clearly shows the dependence of
$\tilde{q}(k) = \tilde{q}(k_h,k_z)$ on the mode-angle with respect to
the vertical as expected by Eq.~(\ref{qg_pvf}) for $f = N$ large.

\begin{figure}
\subfigure[Spectra of horizontal kinetic energy (solid lines) and potential energy (dashed lines)
  as a function of $k_h$ for various fixed $1 \leq k_z \leq 30$ as
  indicated.]{
\includegraphics[scale=.28]{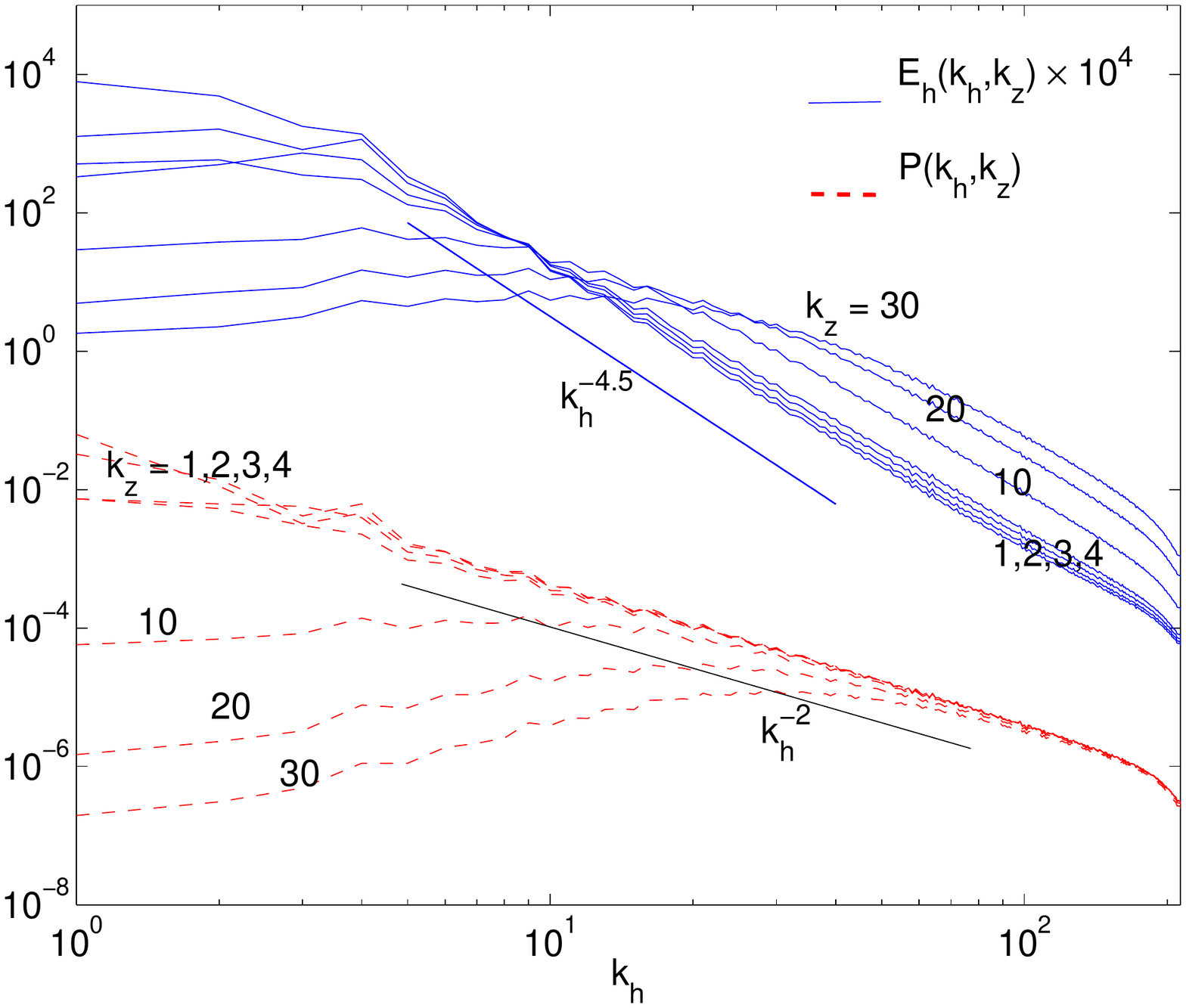}
\label{Ro0Fr0_spec:a}
}~~~~~~
\subfigure[Spectra of horizontal kinetic energy (solid line) and potential energy (dashed line), as functions of $k_h$, each summed over $k_z$.]{
\includegraphics[scale=.28]{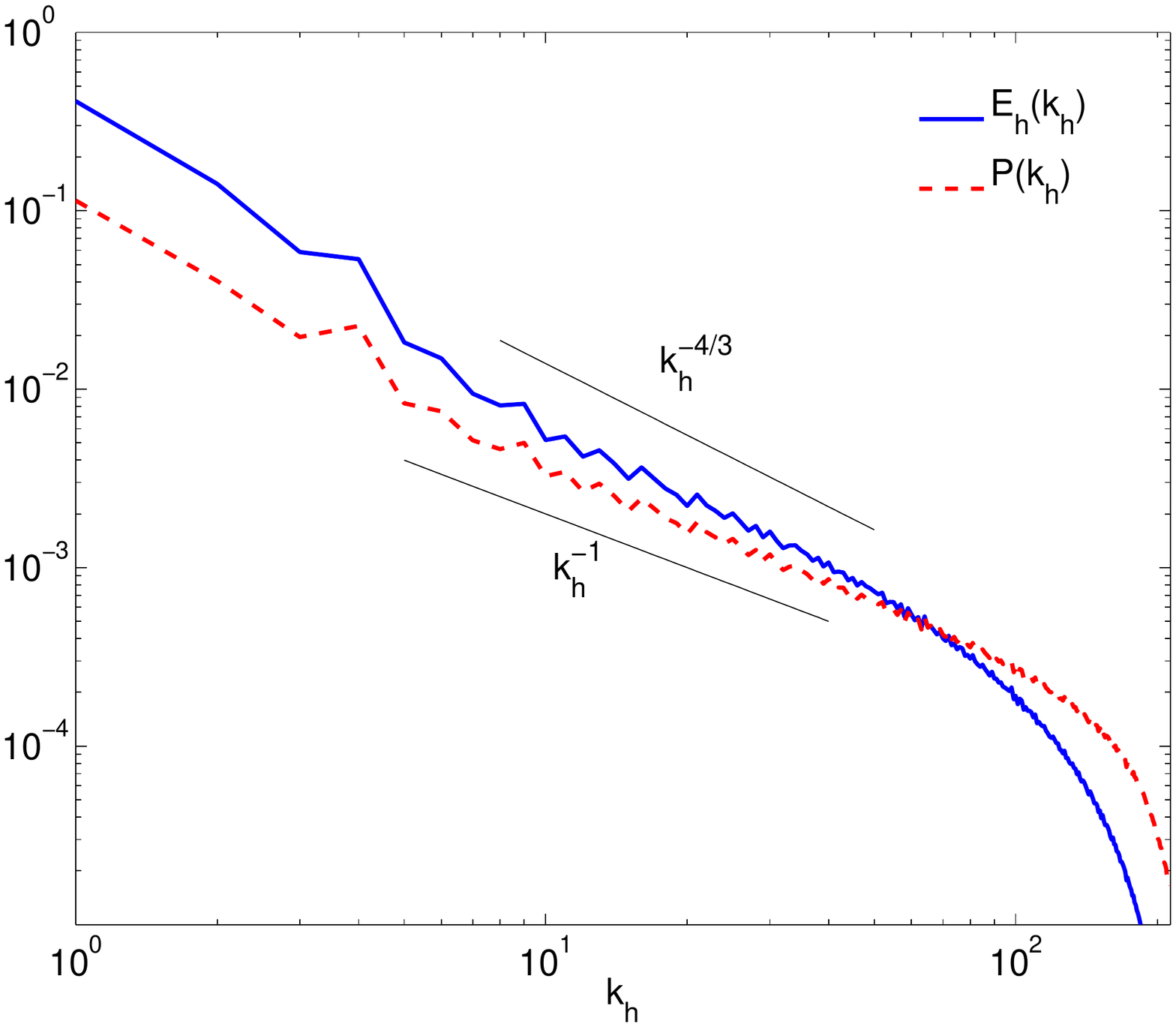}
\label{Ro0Fr0_spec:b}
}
\subfigure[Spectra of horizontal kinetic energy (solid lines) and
potential energy (dashed lines) as a function of $k_z$ for various
fixed $1 \leq k_h \leq 30$ as indicated.]{
\includegraphics[scale=.28]{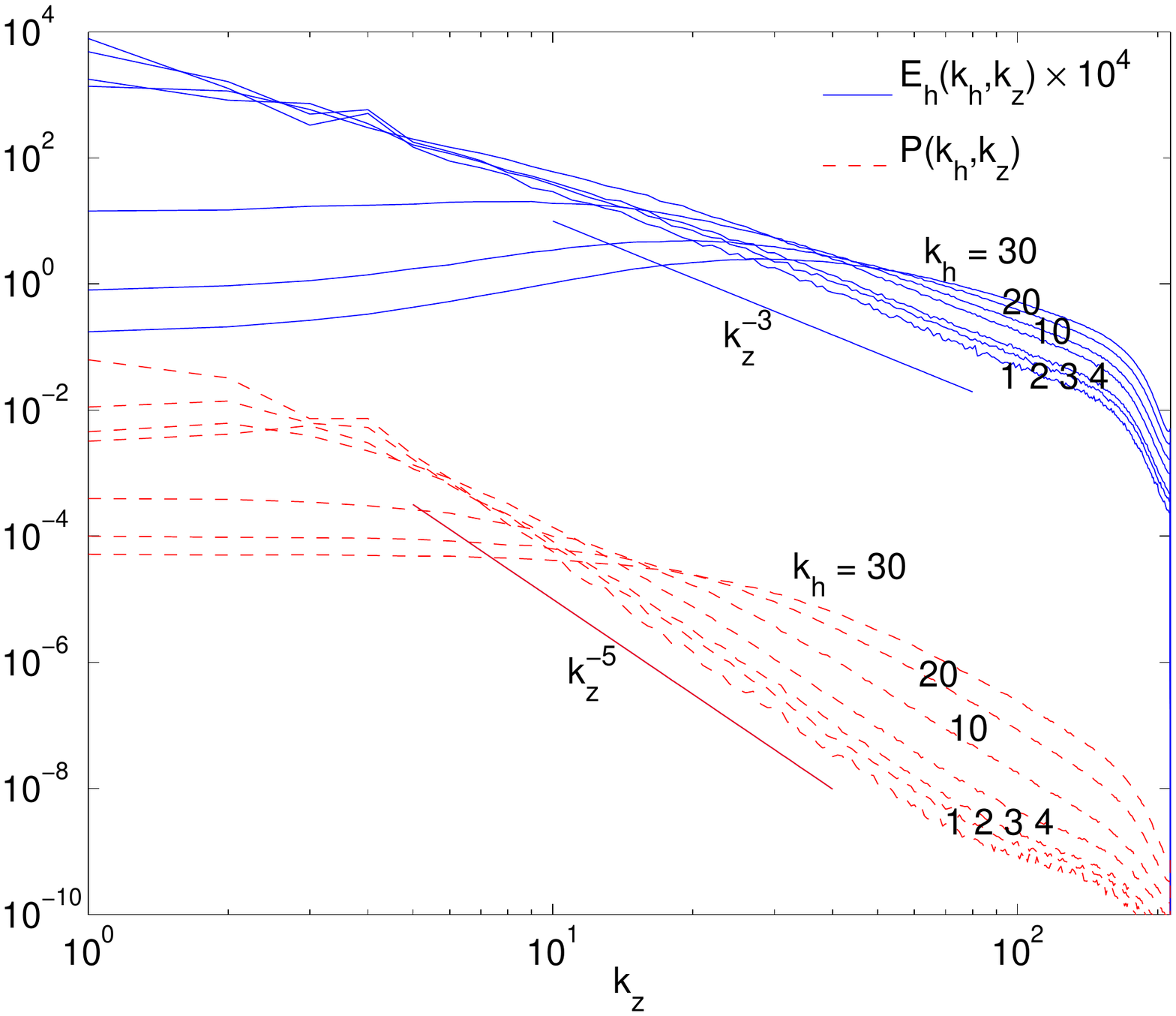}
\label{Ro0Fr0_spec:c}
}~~~~~~
\subfigure[Spectra of horizontal kinetic energy (solid line) and potential energy $P(k_z)$ (dashed line), each summed over $k_h$, as functions of $k_z$.]{
\includegraphics[scale=.28]{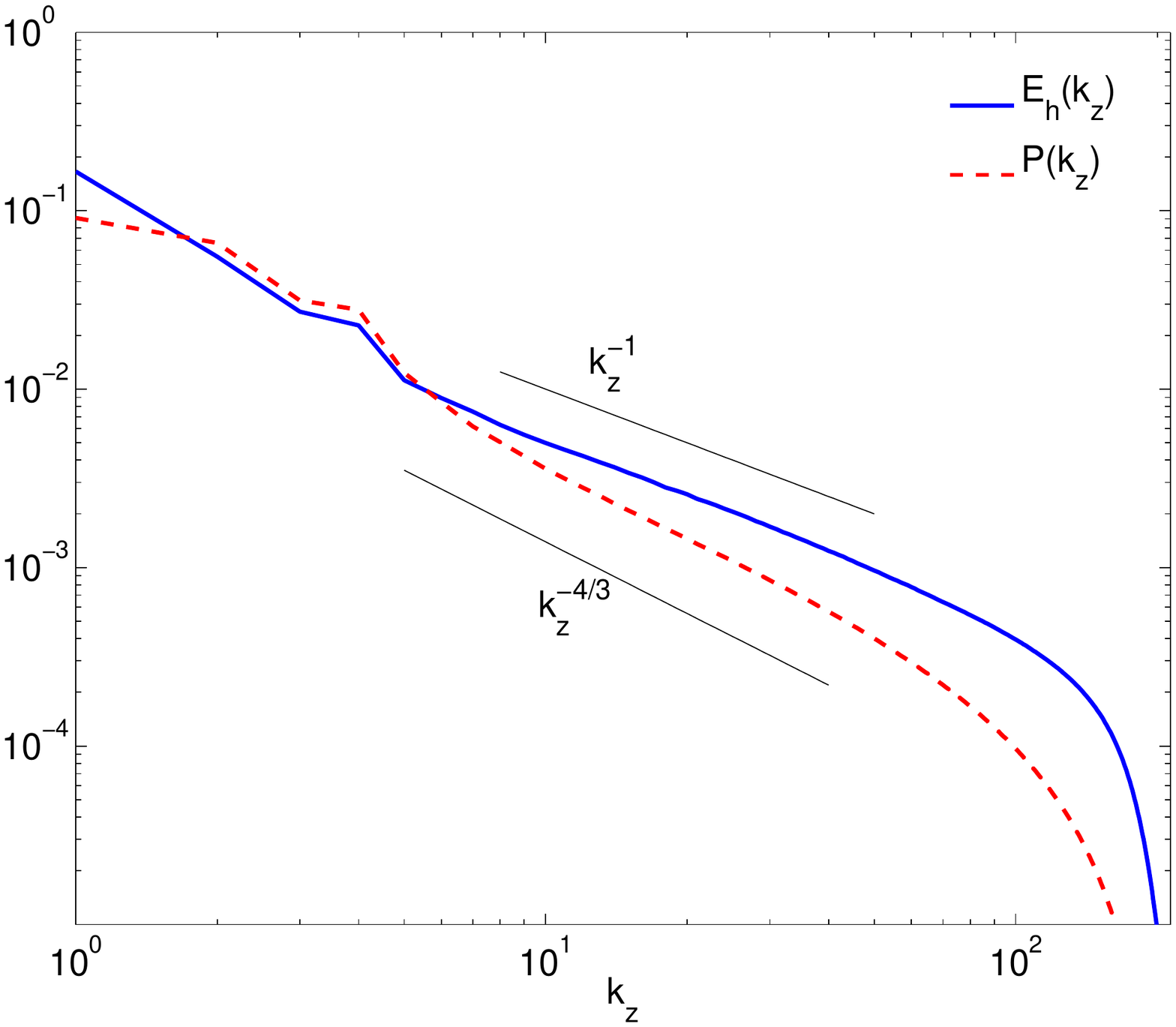}
\label{Ro0Fr0_spec:d}
}
\caption{[Color online] The horizontal kinetic and potential energy spectra for $f=N$
strongly rotating and stratified flow (Table \ref{param_table} data (i)). \label{Ro0Fr0_spec}}

\end{figure}
Next we consider the effect on the scaling of energy spectra as
predicted in Eqs. (\ref{kh-5},\ref{kz-5}). Fig.  \ref{Ro0Fr0_spec:a}
shows $E_h$ and $P$ as a function of $k_h$ for various fixed $k_z$. For
$E_h$, the spectrum for a particular $k_z$ changes over from 
basically flat to slowly increasing for small $k_h$ to close to
$k_h^{-5}$ decay when $k_h > k_z$.  This is in agreement with the
deduction that horizontal kinetic energy is suppressed in the small
aspect ratio modes by the dominance of potential enstrophy in those
modes, and consequently scales according to Eq.(\ref{kh-5}). Note that
there is a cross-over range below which the energy decreases as $k_z$
increases, and above which the energy increases as $k_z$ increases. In
Fig.  \ref{Ro0Fr0_spec:a} the spectra for potential energy for which
we do not have a prediction in this limit of aspect ratio, when
plotted in the same manner as above, also shows aspect-ratio
dependence for $\tau \ll 1$ but with much shallower scaling, the
$k_z^{-2}$ line is shown for comparison. There is no cross-over range
for $P(k_h,k_z)$ and there appears to be much less
suppression of potential energy for modes such that $\tau \ll 1$.

In Fig. \ref{Ro0Fr0_spec:b} we show $E_h(k_h)$ summed over all $k_z$ (top) 
and similarly $P(k_h)$ (bottom). Interestingly, these scale approximately
as $k_h^{-4/3}$ and $k_h^{-1}$ respectively, which is much shallower
than the $k_h^{-5}$ observed or $E_h$ for $\tau \ll 1$, and even
shallower than $k_h^{-5/3}$. Thus, while for a given $k_z$ the
horizontal energy is suppressed for $\tau \ll 1$, the {\it average}
over $k_z$ indicates net transfer to large $k_h$ of both $E_h$ and
$P$. 

To verify the prediction Eq. (\ref{kz-5}) for the large aspect-ratio
modes $\tau \gg 1$ we examine the scaling of $P(k)$ as a function
of $k_z$ for various fixed $k_h$ (Fig. \ref{Ro0Fr0_spec:c}).
Here again the spectrum is nearly a plateau with a turnover to
$k_z^{-5}$ as $\tau \gg 1$. Note again the cross-over range below
which potential energy decreases as $k_h$ increases, and above which
the potential energy increases as $k_h$ increases. $E_h(k)$ plotted
in the same way also shows some aspect ratio dependence but with
much shallower scaling indicating that $E_h$ is not suppressed in the
wavevectors with $\tau \gg 1$. 

In \cite{KurWinTay08} we presented the spectral scaling results for
the case $f=N$ for runs at $512^3$ and all other parameters very
similar, except that in that case only the momentum was forced. In
that paper, we did not have an estimate for steeper than $-3$ scaling
exponent. The resulting spectra behaved very similarly to what we have
presented in Fig. \ref{Ro0Fr0_spec} so far as scaling exponents are
considered. The overall magnitude of the potential energy spectra are
much higher in the scalar-forced case presented here, as is to be
expected. Thus the detailed forcing seems not to affect the results as
far as the scaling exponents go.

To conclude this discussion for the $f=N$ strongly rotating and
stratified case, the suppression of $E_h$ in the small aspect ratio
modes $\bm{k}$ with $\tau \ll 1$ does not result in potential
energy being suppressed in the same way. Conversely, the suppression
of $P$ in the large aspect ratio modes with $\tau \gg 1$, does not
result in the horizontal kinetic energy being suppressed in those
modes.  However, on {\it summing} over all $k_z$ [$k_h$] because of
the strong aspect-ratio dependence, $E_h(k_h)$ [$P(k_z)$] scales with
exponent between $-1$ and ${-4/3}$. Thus a mode-by-mode suppression of
horizontal energy in the small aspect-ratio modes and of the potential
energy in the large aspect-ratio modes, nevertheless results in a net
transfer of these energies into the high wavenumbers {\it on average},
along with the downscale transfer of potential enstrophy. We thus obtain 
{\it a posteriori} justification in our ansatz of $\varepsilon_Q/\varepsilon_d$
dependence (Eqs.  (\ref{kh-5}) and (\ref{kz-5})) for the functional
form of $E_h$ and $P$.

This example also goes to show how one-dimensional spectra in either
the $k_h$ or $k_z$ wave-component may obscure underlying dependencies
on aspect-ratio of the modes and hide the subtle ways by which the
energy populates the high wavenumbers on average. This
possibility was noted in \cite{BGSC06} who studied purely rotating
flow, with no PV, and noticed a dependence of the spectra 
on the mode-angle as well.

\begin{figure}
\centering
\includegraphics[scale=0.45]{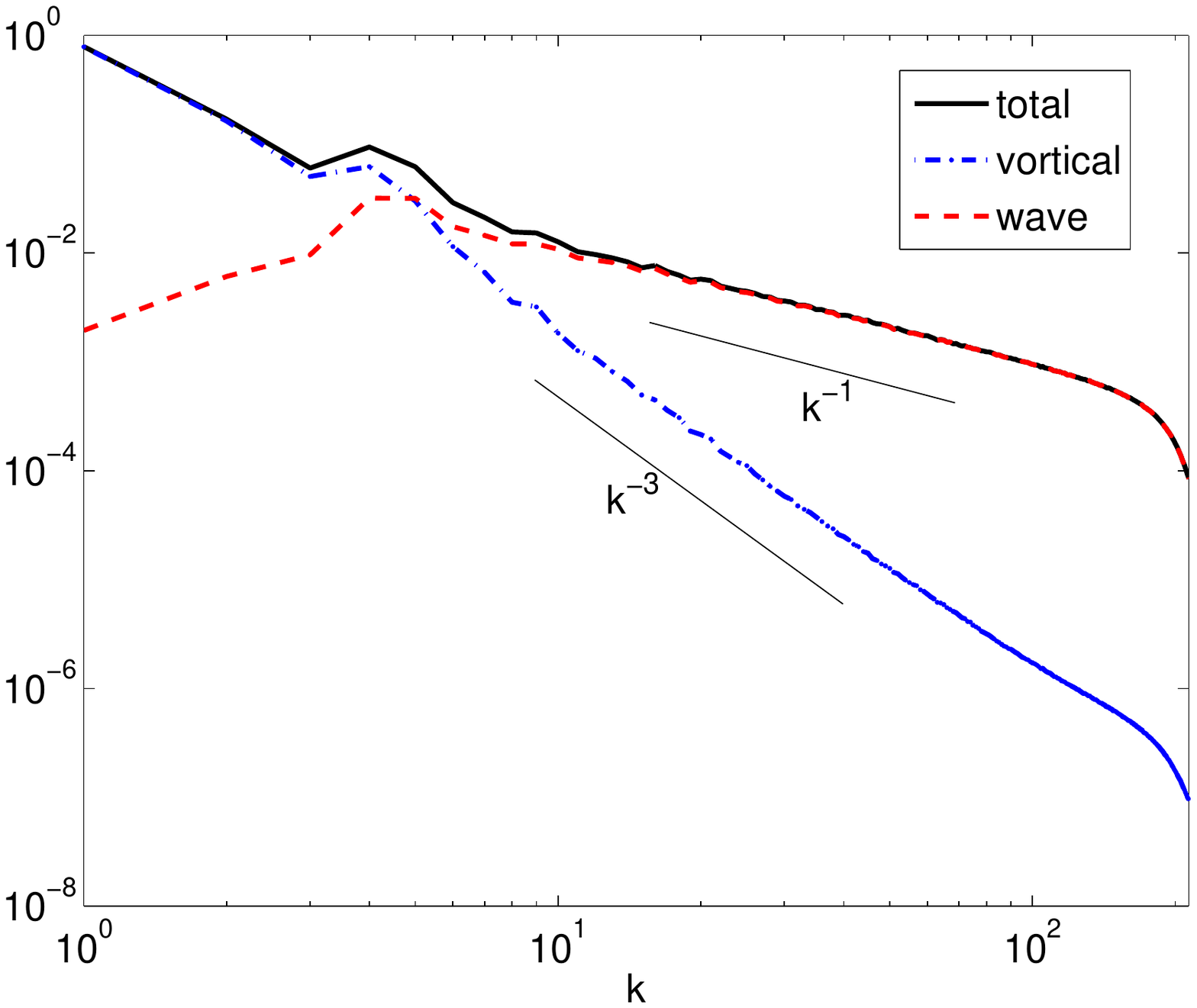}
\caption{[Color online] Wave, vortical and total spectra for run 3 $Ro = Fr = 0.0023$.}
\label{Run3_WVT}
\end{figure}

\section{Flow visualizations}
\begin{figure}[ht]
\subfigure[]{
\includegraphics[scale=0.3]{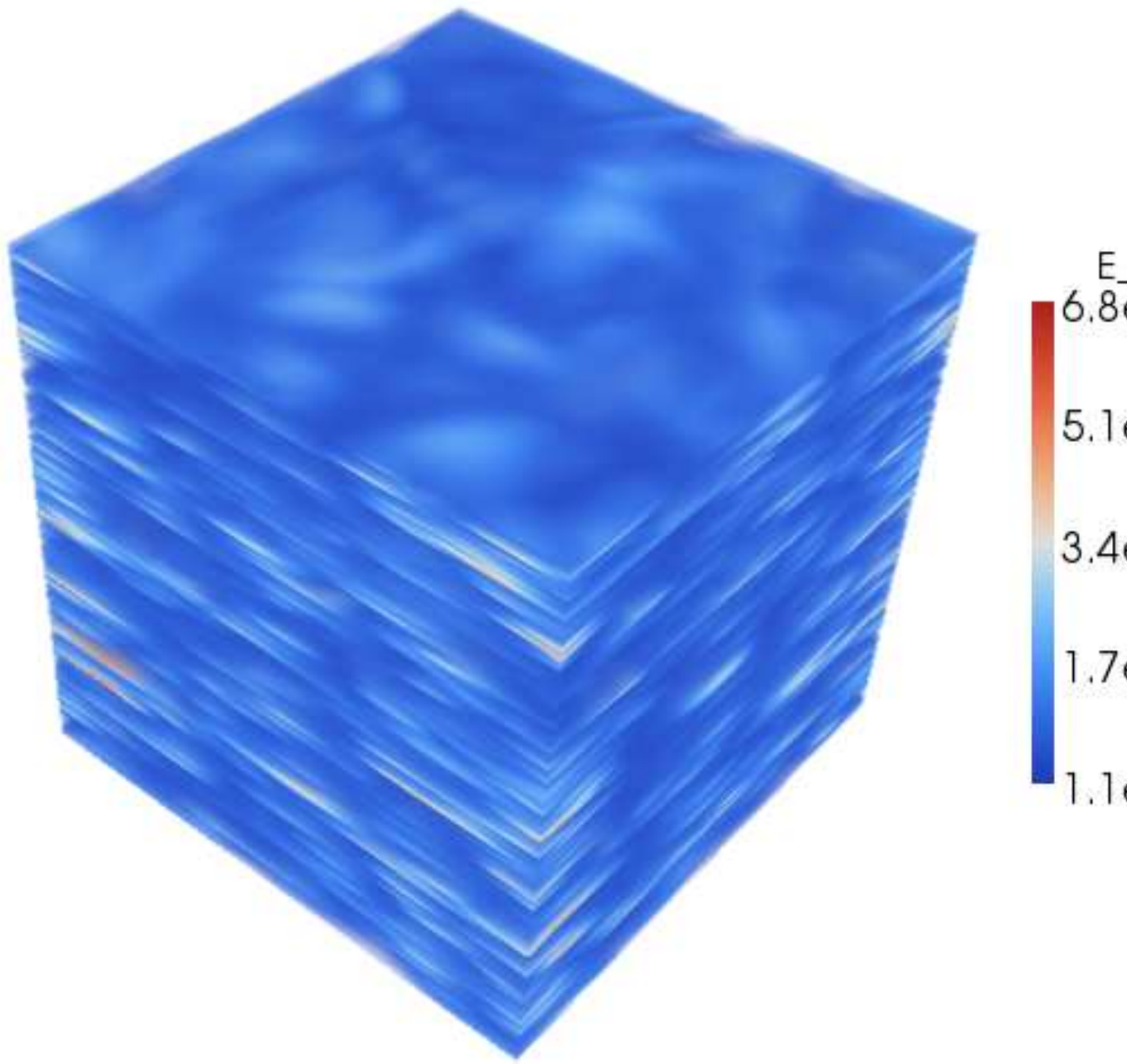}
\label{viz:a}
}
\subfigure[]{
\includegraphics[scale=0.3]{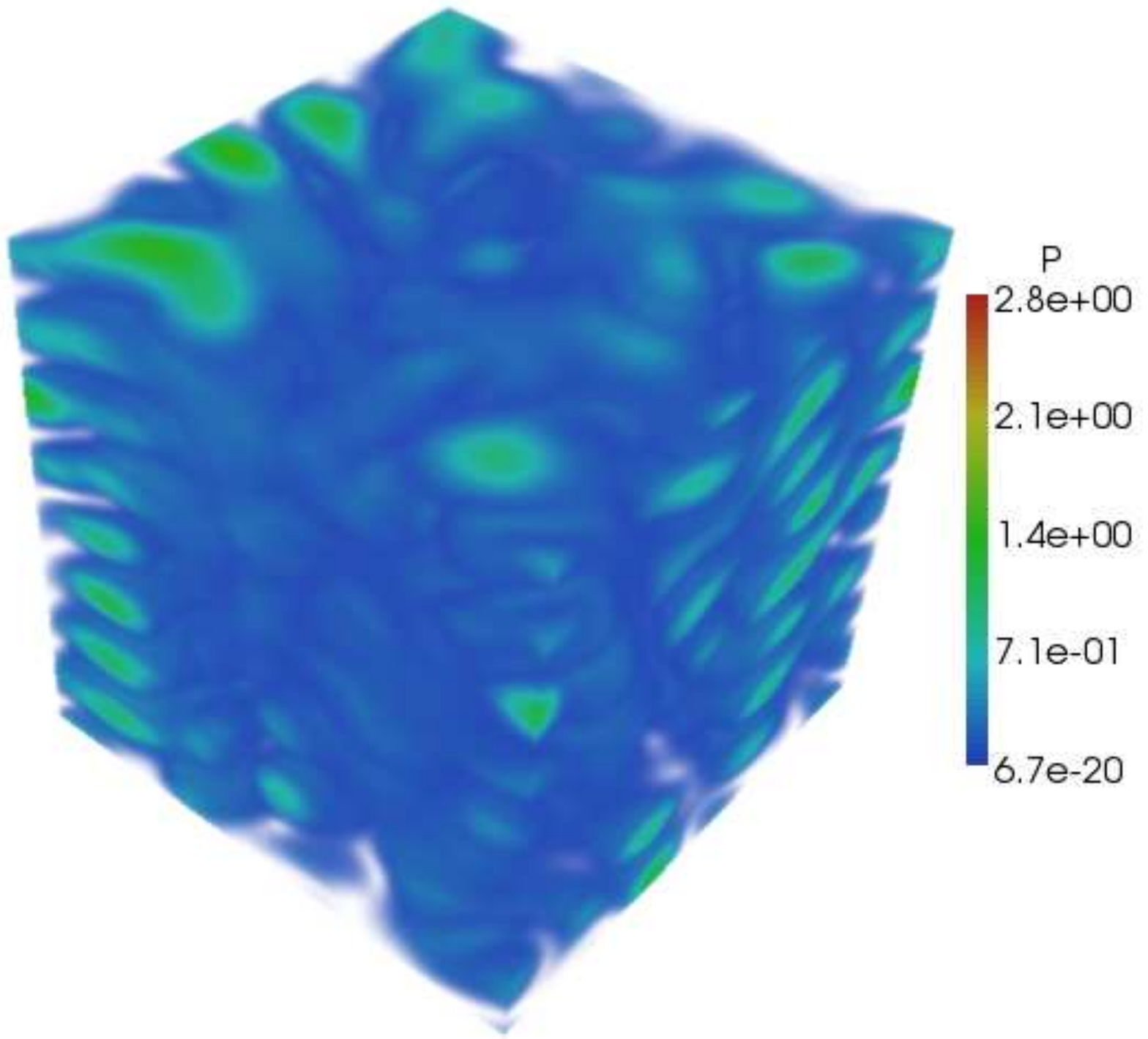}
\label{viz:b}
}
\subfigure[]{
\includegraphics[scale=0.3]{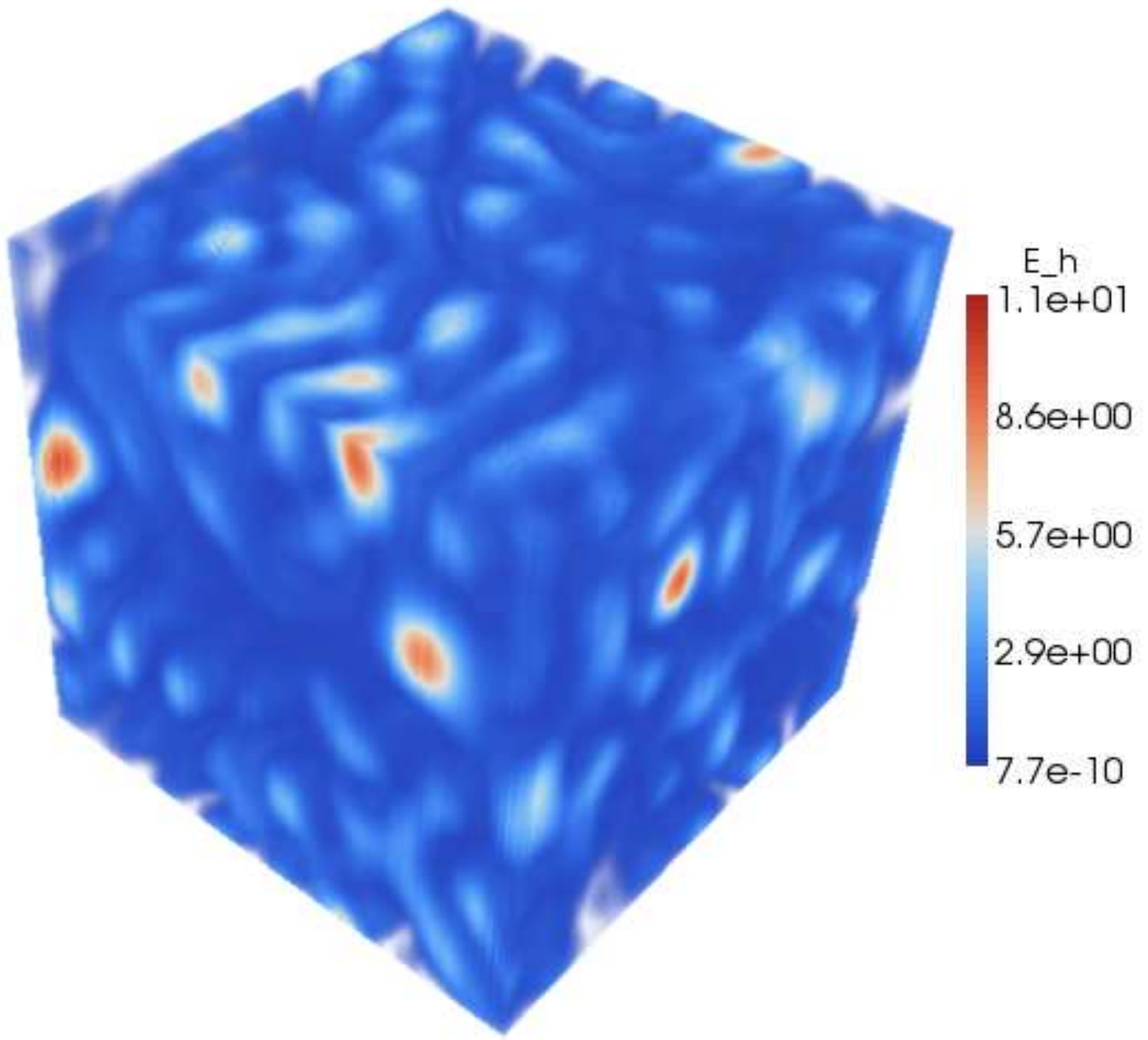}
\label{viz:c}
}
\subfigure[]{
\includegraphics[scale=0.3]{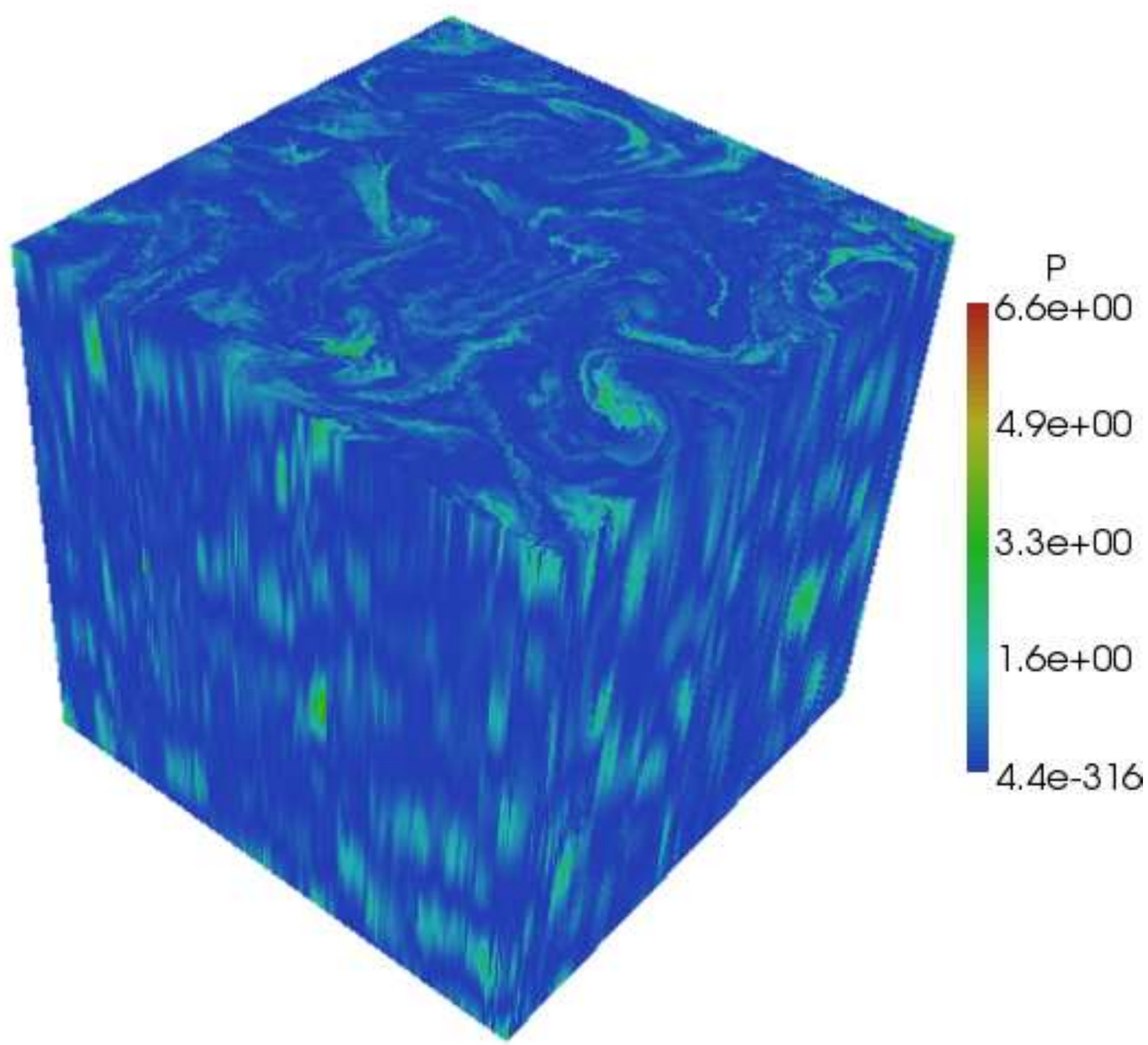}
\label{viz:d}
}
\subfigure[]{
\includegraphics[scale=0.3]{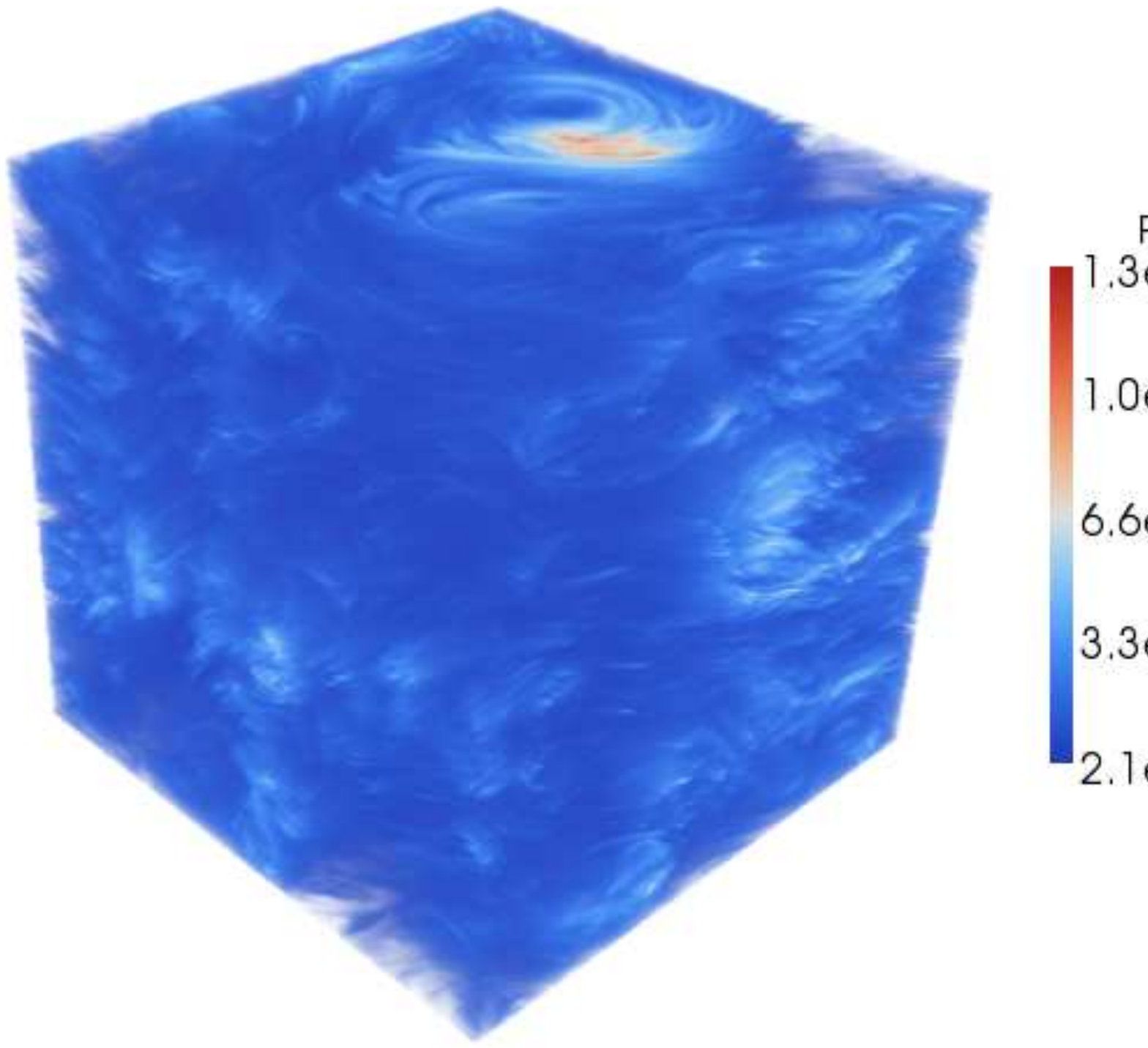}
\label{viz:e}
}
\subfigure[]{
\includegraphics[scale=0.3]{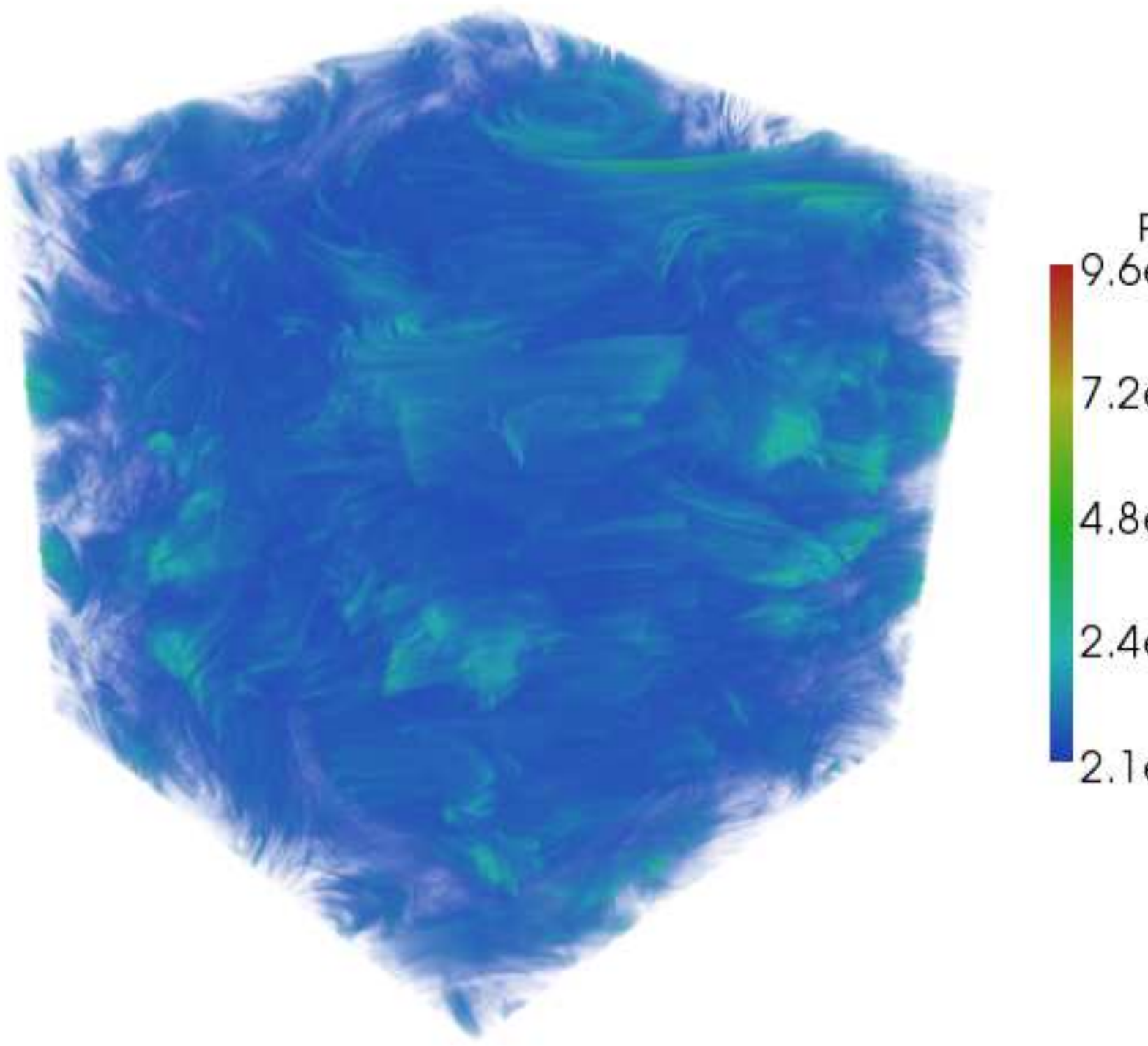}
\label{viz:f}
}
\label{viz}
\caption{[Color online] Volume rendering of horizontal kinetic energy (left) and potential
  energy (right) in the three extreme parameter regimes. Top: run 4, Ro = 1, Fr = 0.001 ; middle: run 5, $Ro=1, Fr = 0.001$; bottom: run 1, $Ro=Fr=0.0023$.}
\end{figure}

To give a qualitative picture of the quantities of interest, we
present volume rendering of snapshots of the fully developed flow for
the horizontal kinetic energy and the potential energy in Fig.
\ref{viz}. It must be noted that it is not clear what correspondence
the spectra have to the structures in the physical space
visualizations of the flow. The spectra are related to the physical
space correlations as fourier transform pairs. A particular
distribution of a field in physical space does not indicate a
unique spectral distribution in fourier. The visualizations presented
here should be treated as qualitative impressions of the flow
structures. Nevertheless, we will point out consistencies between what
is observed in the spectra with what is observed in visualizing the
corresponding physical space quantities.

When stratification dominates over rotation the horizontal kinetic
energy exhibits flat horizontal structures at all values
(Fig.\ref{viz:a}) which appear to have very little extent in the
vertical direction. This is consistent with both Eq. \ref{kh-5} and
the discussion and plots in section \ref{Ro1Fr0} which confirm
complete independence from $k_z$. These structure are quite different
from VSHF since the $k_h=0$ modes have a very small contribution in
this regime. The potential energy for this flow (Fig.  \ref{viz:b})
appears to exhibit some tendency towards horizontally oriented blobs
but retains more three-dimensionally coherent structures at the larger
values, indicating that the potential energy distribution follows the
distribution of horizontal kinetic energy to a degree, consistent with
Fig. \ref{Ro1Fr0_spec:b} but still retains some dependence on $k_z$
for small $k_z$ (large vertical scales) consistent with Fig.
(\ref{Ro1Fr0_spec:d}).

Conversely, when rotation dominates over stratification, Fig.
\ref{viz:d} shows that the potential energy exhibits structures that
are strongly oriented in the vertical direction at all values with
very little extent in the horizontal, consistent with Eq.
(\ref{kz-5}). The horizontal kinetic energy also shows vertically
coherent structures Fig. \ref{viz:c} but also shows high-valued larger
blobs with more horizontal extent showing that $E_h$ also retains some
horizontal scale dependence.

In the case of equal rotation and stratification (Figs. \ref{viz:e},
\ref{viz:f}), horizontal kinetic energy and potential energy look
rather isotropically distributed with no preferred direction, in
agreement with the spectra in Figs. \ref{Ro0Fr0_spec:b} and
\ref{Ro0Fr0_spec:d} which show relatively shallow (ranging between -1
and -4/3) scaling in both $k_z$ and $k_h$ on average.

To summarize, equal rotation and stratification does not show any
preferential direction for the energy isosurfaces, consistent with
nearly the same shallow scaling in $k_h$ and $k_z$ (Figs.
\ref{Ro0Fr0_spec:b}, \ref{Ro0Fr0_spec:d}). Stratification dominated
flows align the energy isosurfaces in the horizontal direction while
rotation dominated flow align the energy isosurfaces along the
vertical. The strongly stratified data are more in agreement with our
asymptotic predictions than the strongly rotating data; the latter
observation could indicate that even stronger rotation is required to
see the asymptotic behavior.

\section{Discussion and Conclusions}
The small scale, high wavenumber spectra of rotating Boussinesq have
never before been studied at such asymptotically low values of $Ro$
and $Fr$ using high-resolution simulations with unbiased isotropic
forcing and dissipation. The rigorous theoretical results for the $Ro
\sim {\cal O}(1), Fr \rightarrow 0$ by \cite{EM98} and the
corresponding simulations of \cite{SW02} extract the $k_h = 0$
wave-modes as the VSHF contribution to the large scales. In our study
of the small scales in this limit, which we have checked have
negligible contribution of the $k_h = 0$ modes (see Fig.
\ref{Run1_WVT}), the vortical modes display a constant spectrum for $k
> k_f$, apparently dominated by the $k_z^0$ behavior of the horizontal
kinetic energy (see Fig. \ref{Ro1Fr0_spec:d}).  This appears to be a
consequence of one of the key new results of this work, namely that
the horizontal kinetic energy becomes independent of (constant in)
$k_z$ for any $k_h$ (see Eq.  \ref{kh-5} and Fig.
\ref{Ro1Fr0_spec:c}) for $\Gamma \ll 1$.  This uniform spectral
distribution implies a very efficient downscale transfer, with energy
contained predominantly by the vortical modes. The exact value of the
steep scaling exponent in $k_h$ is not as important for this result as
the fact that the spectrum becomes independent of (constant in) $k_z$
while decaying sufficiently quickly in $k_h$ such that the spectrum
becomes independent of spherical wavenumber $k$ as well (Fig.
\ref{Run1_WVT}).

The strongly rotating case with finite stratification (runs 2 and 5)
were expected to achieve the limiting behavior of Eq. \ref{kz-5} with
collapse of the spectral curves for all $k_h$ as a function of $k_z$.
While we see the correct trend in this direction, the curves in Fig.
\ref{Ro0Fr1_spec:a} show that the collapse is not complete. We are
thus lead to believe that $Ro$ will need to be driven even lower and
the resolution increased to see the collapse. We can postulate that,
for sufficiently small $Ro$, the potential energy curves for various
$k_h$ will collapse to the same function of $k_z$, corresponding to
Eq. \ref{kz-5}. Then, potential energy as a function of $k$ would be
flat, governed by the flat spectrum in $k_h$ and strongly decaying
spectrum in $k_z$. Again the exact exponent of the decay law would not
affect this general conclusion.

The equally strongly rotating and stratified case (run 3) show
how underlying anisotropic scalings of the spectra, depending on the
cotangent of the mode-angle $\tau$, can be masked by summing over one
or other of the wavevector components, as is commonly done.  In Fig.
\ref{Ro0Fr0_spec:b} we show that the sum over all $k_z$ of the
horizontal kinetic energy $E_h(k_h)$ shallow with scaling $\sim
k_h^{-4/3}$. Thus the {\it average} over all $k_z$ smears out the
effect of suppression by potential enstrophy and results in a much
shallower scaling that the $\tau$ dependent $k_h^{-5}$ scaling. The
net effect is that energy fills the high-wavenumbers with almost the
same shallow scaling in $k_h$ and $k_z$ (Fig. \ref{Ro0Fr0_spec:b} and
\ref{Ro0Fr0_spec:d}). In Fig. \ref{Run3_WVT} the wave-vortical mode
decomposition of the spectra shows that the total downscale energy
scales as $k^{-1}$ and is dominated by wave modes. This is a well-known
result and the scaling is identical to that of the passive scalar
since the wave-modes are passively advected in this limit. What we
further establish from this analysis is that the shallow spectrum is
set up in spite of the underlying steeper, anisotropic,
$\tau$-dependent spectra in $k_h$ and $k_z$.

We have made an attempt at a physically consistent phenomenology that
results in a steeper than -3 scaling of the spectra within this
framework. The $-5$ scaling that arises out of the assumption that the
ratio $\varepsilon_Q/\varepsilon_d$, where $\epsilon_d$ is the net
flux of the energy downscale, plays a role in the downscale dynamics
seems justified {\it a posteriori}. The analysis and verification of
such an ansatz requires a detailed study of the fluxes and dissipation
of both potential enstrophy and energy and is the subject of a
separate work. As we have said before, the main conclusions of this
paper hold irrespective of the specific scaling exponent for the decay
of energy in the $k_h$ and $k_z$.

The prediction for steep decay in $k_h$ according to Eq. (\ref{kh-5})
arises from the constraint that potential enstrophy places on the
horizontal kinetic energy in $k_h$ for $\Gamma \ll 1$. While there is
no explicit constraint on the potential energy, the data analysis in
this regime reveals that potential energy scales very similarly to the
horizontal kinetic energy as a function of $k_h$, indicating that it
too is suppressed in $k_h$. Thus the total energy, where the
contribution from the vertical component of the velocity is measured
to be very small, appears to be suppressed in the large $k_h$ modes.
The observed shallow scaling in $k_z$ of both horizontal kinetic and
potential energies, suggests that the total energy is allowed to
transfer into the large $k_z$ modes.  Similarly, for $\Gamma \gg 1$,
the suppression of potential energy in $k_z$ is mirrored to a large
degree by the horizontal kinetic energy. These observations indicate that the
exchange between kinetic and potential energies is local in
wavenumber in both $f/N \ll 1$ and $f/N \gg 1$ regimes. 

Recent work explored the connection between hyperviscosity, Galerkin
truncation and the energy bottleneck \cite{Frischetal_prl08} in
Navier-Stokes turbulence without rotation or stratification. In that
paper it was proposed that high values of hyperviscosity result in a
partial thermalization of the high wavenumbers which manifests as
energy `bottleneck' phenomenon which contaminates the inertial range.
In the present studies of rotating and stratified turbulence, the only
substantial difference between the hyperviscous and the comparable
runs with regular NS viscosity is the consistently less extended
scaling ranges of the latter runs, which is to be expected.  Therefore
we use the hyperviscous runs to extract scaling exponents while
checking against the NS runs to make sure that bottleneck-type
artifacts are not introduced. It appears that for rotating and
stratified flow and for the value of hyperviscous power used, the
bottleneck-type artifacts, if any are minimal, and do not affect the
purposes of this study.

While we have discussed three extreme parameter regimes, this
discussion encompasses all values of $\Gamma$ thus
allowing us to analyze arbitrary $f/N$ in the limit of linear (or near
linear) PV. In an idealized infinite domain where $k \rightarrow
\infty$, for any fixed $f/N$ the $\Gamma \ll 1$ regime is achieved
when $k_h/k_z \gg f/N$ while the $\Gamma \gg 1$ regime is achieved
when $k_h/k_z \ll f/N$. The transition between regimes occurs for
$\Gamma = 1$ or when $1/\tau = k_h/k_z = f/N$. Thus we observe that
this type of analysis could be very relevant to geophysical flow
regimes where $f$ and $N$ values are not as extreme but where the
scaling ranges are much longer.

\acknowledgements The author is grateful for the hospitality of the
Courant Institute (Climate Atmosphere Ocean Science program) for a
year-long visit in 2008-2009 during which much of the work for this
paper was completed. The author thanks Leslie Smith for numerous
valuable discussions related to this work and for comments on an early
draft of the paper. Early comments and suggestions by Oliver Buhler,
Andrew Majda and Shafer Smith are gratefully acknowledged. Mark Taylor
provided generous assistance with the Boussinesq code and its optimal
scaling on the Argonne BG/P machine. This research used resources of
the Argonne Leadership Computing Facility at Argonne National
Laboratory, which is supported by the Office of Science of the U.S.
Department of Energy under contract DE- AC02-06CH11357. Partial
funding was provided by the DOE Office of Science, ASCR program in
Applied Mathematics.

\bibliographystyle{jfm}
\bibliography{/home/skurien/papers/Bibs/kurien}

\end{document}